\documentclass[11pt]{article}

\RequirePackage{amsthm,amsmath,amsfonts,amssymb}
\RequirePackage[authoryear]{natbib}
\RequirePackage[colorlinks,citecolor=blue,urlcolor=blue]{hyperref}
\RequirePackage{graphicx}
\RequirePackage[percent]{overpic}
\RequirePackage{mathtools}
\RequirePackage{bm}
\RequirePackage{booktabs}
\RequirePackage{array}
\RequirePackage{longtable}
\RequirePackage{tabularx}
\RequirePackage{makecell}
\RequirePackage{xcolor}
\RequirePackage{enumitem}
\RequirePackage{placeins}

\newcolumntype{Y}{>{\raggedright\arraybackslash}X}

\def\ba{{\bm a}}

\def\bx{{\bm x}}

\def\bA{{\bm A}}
\def\bP{{\bm P}}
\def\bX{{\bm X}}
\def\bc{{\bm c}}

\newcommand{\bbeta}{\mbox{\boldmath $\beta$}}

\newcommand{\btheta}{\mbox{\boldmath $\theta$}}

\newcommand{\bmu}{\mbox{\boldmath $\mu$}}
\newcommand{\brho}{\mbox{\boldmath $\rho$}}
\newcommand{\etam}{\mbox{\boldmath $\eta$}}

\begin{document}

\title{Linked-Tucker Factorized Individualized Regression for Paired Multivariate Categorical Outcomes}
\author{Arkaprava Roy$^{1}$, Jeremy T. Gaskins$^{2}$, Steven Levy$^{3,4}$,\\
and Somnath Datta$^{1}$\\[1ex]
\small $^{1}$Department of Biostatistics, University of Florida\\
\small $^{2}$Department of Bioinformatics \& Biostatistics, University of Louisville\\
\small $^{3}$Department of Preventive and Community Dentistry, University of Iowa\\
\small $^{4}$Department of Epidemiology, University of Iowa}
\date{}
\maketitle

\begin{abstract}
We propose a joint individualized hurdle-ordinal regression model for paired zero-inflated ordinal outcomes with subject-specific, spatially-varying, and time-varying covariate effects, motivated by the Iowa Fluoride Study (IFS). The two outcomes of interest, dental caries and dental fluorosis, are each measured repeatedly over multiple ages at fine spatial resolution, across tooth surfaces for caries and tooth zones for fluorosis, yielding a nested longitudinal structure with substantial zero inflation, ordinality, and heterogeneity across individuals and locations. For each outcome, a hurdle component models whether the observation crosses from disease absence to disease presence, while a proportional-odds severity component models the ordered disease level among positive observations. To represent the high-dimensional regression coefficient arrays parsimoniously, we introduce a linked Tucker tensor factorization. Shared subject-mode factors across the two Tucker decompositions induce dependence between the caries and fluorosis coefficient tensors, while separate spatial factors accommodate the distinct measurement grids
of the two outcomes.
A horseshoe prior on the core tensor elements encourages sparsity, and posterior computation is carried out via the No-U-Turn Sampler implemented in NumPyro. Population-level effect summaries are obtained by projecting individualized posterior linear predictors onto the design space, and Wasserstein barycenters aggregate these summaries across tooth locations and anatomical classes. Applied to the IFS, the model reveals component-specific and spatially heterogeneous associations between early-life fluoride and dietary exposures and both outcomes: fluoride exposure is associated with both increased odds of developing fluorosis and greater fluorosis severity among affected teeth, while soft drink intake consistently increases the probability of developing caries. These patterns, which differ between presence and severity components and vary across tooth locations, ages, and sub-populations defined by prior caries status, underscore the importance of the joint hurdle-ordinal framework for disentangling disease occurrence from disease progression in complex multilevel dental data.
\end{abstract}


\section{\label{sec:intro} Introduction}

Many medical, dental, and public health studies generate outcomes that are simultaneously
longitudinal, spatially-structured, ordinal, and zero-inflated. Repeated
measurements are often collected over time, at multiple locations within the
same individual, and on ordered disease scales. Consequently, the resulting
responses exhibit within-subject dependence across time and space together with
substantial heterogeneity across individuals. These features create major
challenges for statistical modeling and interpretation, particularly when the
scientific goal is to identify both risk-increasing and protective covariates
and to understand how their effects evolve over time.

This work is motivated by the Iowa Fluoride Study (IFS), a long-term cohort
study designed to understand how fluoride exposures and behavioral factors shape
dental health from childhood to early adulthood \citep{warren2001prevalence,levy2001patterns,Levy2003,Marshall2003,marshall2005diet,marshall2005roles,warren2002dental,warren2021measurement,curtis2020decline,yazdanbakhsh2024community}. The two outcomes of primary
interest are dental caries and dental fluorosis. Dental caries is a disease process marked by areas of tooth decay that can progress to cavitation, whereas dental fluorosis is a
developmental enamel condition associated with excessive fluoride intake
during enamel formation. In the IFS, both outcomes are measured repeatedly over
multiple ages and are recorded at fine spatial resolution. For each
participant, caries is observed across tooth surfaces and fluorosis across
tooth zones, yielding a nested longitudinal structure with surfaces or zones
within teeth, teeth within individuals, and repeated observations over time.
The scientific objective is not only to identify risk and protective factors,
such as fluoride exposures and dietary habits, but also to understand how their
effects vary across individuals, tooth locations, and developmental periods
\citep{Levy2003,Marshall2003,Broffitt2013}.

Two features of these responses are especially important: zero inflation and
ordinality. Many tooth-location measurements are disease free, leading to a
large concentration of zeros \citep{Lambert1992,Bohning1999}, while affected observations are recorded on
ordered categorical scales reflecting increasing severity. A useful model must
therefore separate disease absence from disease severity \citep{Rose2006} while also allowing
for spatial, temporal, and subject-specific heterogeneity. In addition,
regression effects may vary across individuals and locations and may change as
teeth develop and exposures accumulate. Standard pooled regression models are
not flexible enough to accommodate these interacting sources of variability.

Existing methods address only parts of this problem. Generalized linear mixed
models incorporate random effects to account for dependence \citep{Zeger1992,Zhang2012}, but the resulting
coefficients are typically conditional on latent random effects and can be
hard to interpret marginally \citep{Neuhaus1991,Gardiner2009}. Generalized estimating equations provide a
population-averaged interpretation \citep{Zeger1992}, but do not specify a full likelihood and
are less convenient for the joint modeling of multilevel zero-inflated ordinal
responses. Copula-based approaches allow flexible dependence modeling across
outcomes \citep{Sklar1959,Pitt2006,Kolev2009,Smith2012}, but the existing literature has largely focused on count outcomes or
simpler spatial structures \citep{mukherjee2024modeling}. Bayesian hierarchical models for dental data have been proposed
\citep{Choo-Wosoba2016,Choo-Wosoba2018,Kang2021,Kang2023}, but these address only one outcome at a time and do not
jointly model the paired caries--fluorosis structure. To our knowledge, there is not yet a unified
framework for jointly modeling caries and fluorosis as paired zero-inflated
ordinal outcomes while allowing subject-specific covariate effects that vary
across space and time.

In this paper, we propose an individualized hurdle-ordinal regression model
for paired zero-inflated ordinal outcomes with spatial and longitudinal
structure. For each outcome, the first component models whether the observation
crosses the hurdle from disease absence to disease presence, while the second
component models the ordinal severity among positive outcomes. This
decomposition distinguishes covariate effects on disease occurrence from
covariate effects on disease severity, which is scientifically important in the
IFS because fluoride may protect against caries while increasing fluorosis
risk.

To capture heterogeneity and dependence parsimoniously, we represent the
regression coefficient arrays through a linked Tucker tensor factorization.
Rather than estimating unrelated coefficients for each subject, location,
predictor, time point, and outcome, we model the coefficient tensors through
low-rank latent factors. This induces structured dependence across modes,
borrows information across sparsely observed cells, and yields a coherent joint
model for caries and fluorosis. A shared subject factor links the two outcomes,
allowing common latent susceptibility profiles to drive dependence between the
paired dental processes. Because caries surfaces and fluorosis zones are not
measured on identical spatial grids, we do not force a shared spatial factor
across the two outcomes. The outcomes nevertheless share tooth-level anatomy;
in the present model, this dependence is captured indirectly through the shared
subject factors rather than by imposing a common spatial factor on two
non-identical location grids.

The proposed framework provides both individualized and marginal summaries.
Posterior draws of location-specific linear predictors are projected onto a
common covariate space to obtain interpretable population-level effect
summaries, while the underlying tensor structure preserves subject-level and
location-level heterogeneity. The model also handles incomplete outcome data
naturally under an ignorability assumption: inference is based on the observed-data likelihood, and the low-rank factorization borrows strength across nearby
subjects, locations, predictors, and time points without requiring explicit
imputation of unobserved responses.

In summary, our framework has five main advantages. First, it provides a direct
separation between disease occurrence and disease severity. Second, it jointly
models caries and fluorosis within a single probabilistic framework. Third, it
accommodates zero inflation, ordinality, spatial clustering, and longitudinal
dependence simultaneously. Fourth, it allows rich subject-specific
heterogeneity while still producing interpretable marginal summaries. Fifth,
its low-rank tensor structure stabilizes estimation in sparse tooth-location
cells by borrowing information across modes.
Although motivated by the Iowa Fluoride Study, the proposed methodology applies
more broadly to studies involving zero-inflated ordinal outcomes with repeated
measurements and spatial structure. 

\section{\label{sec:variables} Study Design and Variables}
The Iowa Fluoride Study (IFS) is a prospective cohort study that enrolled families of newborns in Iowa in 1992--1995 and has followed participants through early adulthood \citep{Levy2003,Marshall2003,Broffitt2013,Marshall2021,Zafar2022}. Dental examinations and questionnaire-based exposure assessments were conducted at multiple ages. The analyses presented here use two data configurations: (i) a cross-sectional analysis of dental caries at age 5 and dental fluorosis at age 9, regressed on covariates assessed from birth to age 5; and (ii) a longitudinal analysis of caries and fluorosis at ages 9, 13, and 17, regressed on time-varying covariates assessed in the periods preceding each examination.

\paragraph*{Outcome variables}
\emph{Dental caries} is recorded for each tooth surface. At age 5, the teeth present are primary (baby) teeth. At ages 9, 13, and 17, the dentition transitions to permanent teeth, with increasing numbers of permanent surfaces available at each exam. The caries scale at each surface ranges from 0 (sound) to 4, with higher categories representing increasing levels of decay severity.
Surfaces with no observed decay contribute the large proportion of zeros.

\emph{Dental fluorosis} is recorded for each tooth zone (cervical [C], incisal [I], middle [M], and occlusal [O]) rather than at surfaces. Fluorosis develops during enamel formation in early childhood; by age 9 the permanent incisors and first molars are fully erupted and assessable. The fluorosis scale at each zone ranges from 0 (no fluorosis) to 3, with higher categories representing increasing levels of enamel change severity. Because fluorosis is a developmental condition, its severity at a given zone reflects cumulative fluoride intake during the enamel formation window for that tooth, which varies by tooth type.

\paragraph*{Predictor variables}
All predictors were assessed by questionnaires
completed approximately every six months
and are summarized for each analysis period as described below. In the cross-sectional (age-5/age-9) analysis, all predictors reflect the period from birth to age 5. In the longitudinal analysis, predictors are time-varying and reflect the multi-year period preceding each examination.

\begin{itemize}[noitemsep]
\item {Dental exam}: Indicator for whether the child had a non-study dental examination in the relevant period. This serves as a proxy for dental care access and health-seeking behavior.
\item {Fluoride intake}: Area under the curve (AUC) of estimated daily fluoride intake in milligrams, integrated over the measurement period. This captures cumulative systemic fluoride intake from water, dietary fluoride supplements, and diet. The AUC formulation accounts for the full intake trajectory rather than a single time point.
\item {Soft-drink intake}: AUC of reported daily soft drink consumption in fluid ounces, integrated over the measurement period. Soft drinks are a primary dietary source of fermentable carbohydrates associated with caries risk.
\item {Tooth brushing frequency}: Average number of times per day teeth were brushed, reported by the parent or child. This reflects oral hygiene behavior.
\item \emph{Non-study dental visits in the past 6 months}: Average number of
non-study dental visits in the 6 months preceding each questionnaire. This
variable measures the frequency of dental-care utilization, whereas the
dental-exam indicator records whether a dental examination occurred. Although
the two predictors may be correlated, they represent distinct aspects of
dental care and are therefore retained in the joint model. Accordingly, their
estimated effects should be interpreted as conditional associations, with
each coefficient describing its association after adjustment for the other.
\item {Fluoride treatments in past 6 months}: Average number of professional fluoride treatments received in the 6 months preceding each questionnaire. This reflects clinical preventive care.
\item {Home water fluoride}: Average fluoride concentration (parts per million) of the household's primary drinking water source over the relevant analysis period. This captures ongoing background fluoride exposure from water.
\end{itemize}

\paragraph*{Tooth surfaces and zones}
For caries, tooth surfaces are coded as buccal (b), distal (d), lingual (l), mesial (m), occlusal (o), occluso-distal (od), and occluso-mesial (om). Not all surfaces are present on all tooth types; for example, the occlusal surface applies to posterior teeth only. For fluorosis, the cervical (C), incisal (I), and middle (M) zones are regions of the buccal surface, while the occlusal (O) zone is assessed on posterior teeth. Tooth types are classified as incisor, canine, premolar, and molar.

\paragraph*{Note on the number of assessable teeth}
The number of teeth and surfaces available for examination varies by age. At age 5, only primary teeth are present. At ages 9, 13, and 17, the dentition is increasingly permanent, with different tooth types erupting at different developmental stages. The model accommodates this naturally through the observed-data likelihood: only observed tooth-surface or tooth-zone combinations contribute to inference at each time point, and the low-rank Tucker structure borrows information across subjects and locations even when some cells are unobserved.

\section{\label{sec:tensor_decomp} Overview of Tensor Decompositions}

This section briefly summarizes the two widely used tensor
factorization frameworks \citep{hitchcock1927expression,tucker:1966,de_lathauwer_etal:2000,kolda2009tensor}.
Consider a $d_{1} \times \dots \times d_{p}$ tensor
$\btheta = \{\theta_{h_{1},\dots,h_{p}}: h_{j}=1,\dots,d_{j},\; j=1,\dots,p\}$.
A common approach for representing such a tensor is through a
low-rank factorization.

\paragraph*{Canonical Polyadic (CP) Decomposition}

The tensor $\btheta$ is said to admit a CP (also known as PARAFAC,
or Parallel Factor Analysis)
decomposition with rank $r$ if its entries can be written as
\[
\theta_{h_{1},\dots,h_{p}}
=
\sum_{z=1}^{r}
\eta_{z}
\prod_{j=1}^{p}
a_{z}^{(j)}(h_{j}),
\qquad
\text{for each } (h_{1},\dots,h_{p}).
\]

Here $\ba_{z}^{(j)} =
[a_{z}^{(j)}(1),\dots,a_{z}^{(j)}(d_{j})]^{\top}$,
for $z=1,\dots,r$ and $j=1,\dots,p$,
are vectors of length $d_j$.
This representation substantially reduces the number of free parameters,
from $\prod_{j=1}^{p} d_j$ in the full tensor
to $r\sum_{j=1}^{p} d_j$ after factorization.

Equivalently, the CP representation can be written compactly as
\[
\btheta
=
\sum_{z=1}^{r}
\eta_{z}
\,\ba_{z}^{(1)} \circ \cdots \circ \ba_{z}^{(p)},
\]
where $\circ$ denotes the vector outer product.

\paragraph*{Tucker Decomposition}

An alternative representation is provided by the Tucker decomposition,
which generalizes the CP model by introducing a core tensor.
Specifically, $\btheta$ admits a Tucker factorization
with multilinear rank $(r_{1},\dots,r_{p})$ if
\[
\theta_{h_{1},\dots,h_{p}}
=
\sum_{z_{1}=1}^{r_{1}} \cdots
\sum_{z_{p}=1}^{r_{p}}
\eta_{z_{1},\dots,z_{p}}
\prod_{j=1}^{p}
a_{z_{j}}^{(j)}(h_{j}),
\qquad
\text{for each } (h_{1},\dots,h_{p}).
\]

The coefficients
$\etam = \{\eta_{z_{1},\dots,z_{p}}\}$
form an $r_{1}\times\dots\times r_{p}$ core tensor,
where $1\le r_j \le d_j$.
The matrices
$\bA^{(j)} = [\ba_{1}^{(j)},\dots,\ba_{r_j}^{(j)}]$
are $d_j \times r_j$ factor matrices with columns
$\ba_{z_j}^{(j)} =
[a_{z_j}^{(j)}(1),\dots,a_{z_j}^{(j)}(d_j)]^{\top}$.

Under this representation,
the number of parameters reduces to
\[
\prod_{j=1}^{p} r_j
+
\sum_{j=1}^{p} r_j d_j
\approx
\prod_{j=1}^{p} r_j,
\]
which can be dramatically smaller than $\prod_{j=1}^{p} d_j$
when the core tensor is low-dimensional.

The Tucker model can also be written using tensor-matrix products:
\[
\btheta
=
\etam
\times_1 \bA^{(1)}
\cdots
\times_p \bA^{(p)},
\]
where $\times_j$ denotes multiplication along the $j$-th mode.

\paragraph*{Relationship between CP and Tucker models}

The CP representation can be viewed as a special case of the Tucker
decomposition obtained when the core tensor is diagonal,
i.e.,
\[
\eta_{z_{1},\dots,z_{p}}
=
\eta_{z}\,
1\{z_{1}=\dots=z_{p}=z\},
\]
with $r_1=\dots=r_p=r$.

Conversely, a Tucker model can be expressed as a CP representation
with rank $r=\prod_{j=1}^{p} r_j$ by vectorizing the core tensor and
appropriately repeating entries in the factor matrices.
Thus, the two formulations are closely related but differ in their
parameterization and degree of compression.

Additional background on tensor factorizations and their statistical
applications, including Bayesian approaches, can be found in
\cite{guhaniyogi2020bayesian,shi2023tensor}.

\section{\label{sec:model} Paired Hurdle-Ordinal Regression with Linked Tucker Factorization}

In our application, the coefficient arrays vary across subject, spatial
location, predictor, and in the longitudinal setting, time. It is therefore
natural to view them as tensors. Tucker factorization provides a parsimonious
representation of such high-dimensional coefficient arrays by decomposing them
into a low-dimensional core tensor and a collection of mode-specific factor
matrices.

Let $C^{i}_{q}$ denote the caries response for subject $i$ at caries location
$q\in\mathcal Q_C$, where $q$ indexes a specific $(\text{tooth},\text{surface})$
pair. Let $F^{i}_{q'}$ denote the fluorosis response for subject $i$ at
fluorosis location $q'\in\mathcal Q_F$, where $q'$ indexes a specific
$(\text{tooth},\text{zone})$ pair. The spatial index sets $\mathcal Q_C$ and
$\mathcal Q_F$ are generally different because tooth surfaces and tooth zones
are distinct measurement units. Thus, for each subject we observe two paired
ordinal outcomes on related but non-identical spatial supports.

\subsection{\label{subsec:hurdle} Individualized hurdle-ordinal model}

Both $C^{i}_{q}$ and $F^{i}_{q'}$ exhibit a large proportion of zeros. We
therefore use a hurdle model \citep{Rose2006,Kang2021,Kang2023} rather than a standard ordinal regression. The
hurdle component models whether the response is zero or positive, while the
ordinal component models severity among the positive categories only. This
avoids the incompatibility that arises when the cumulative probability formula
is written starting at the zero class.

\paragraph*{Caries model}
Let $C^{i}_{q}\in\{0,1,\ldots,C-1\}$, where $0$ denotes the disease-free
category and $1,\ldots,C-1$ denote increasing positive severities. We allow
the predictor sets entering the hurdle and severity components to differ. Let
$\bx^{(o)}_{i}$ denote the predictor vector for the 
occurrence (hurdle)
component and
$\bx^{(s)}_{i}$ the predictor vector for the severity component. We include an
intercept in $\bx^{(o)}_{i}$ for the occurrence model but exclude it from
$\bx^{(s)}_{i}$ because the proportional-odds cutpoints already provide
category-specific intercepts.
The hurdle or occurrence
component models the \emph{probability of
developing caries} (i.e., crossing from disease-free to disease-present):
\[
1 - \pi^{(C)}_{iq}
:=
P(C^{i}_{q}>0\mid \bx^{(o)}_i)
=
\operatorname{logit}^{-1}\!\bigl((\bmu_{i,q})^{\top}\bx^{(o)}_i\bigr),
\]
so that $\pi^{(C)}_{iq} = 1 - P(C^{i}_{q}>0\mid \bx^{(o)}_i)$ is the
probability of remaining caries-free. A positive coefficient in $\bmu_{i,q}$
therefore indicates \emph{increased odds of developing caries}, the natural
direction for a disease-occurrence model.

Conditional on $C^{i}_{q}>0$, we model the positive categories through a
proportional-odds model,
\[
P(C^{i}_{q}\le u\mid C^{i}_{q}>0,\bx^{(s)}_i)
=
\operatorname{logit}^{-1}\!\bigl(\alpha_u-(\bbeta_{i,q})^{\top}\bx^{(s)}_i\bigr),
\qquad u=1,\ldots,C-2.
\]
In this severity component, positive coefficients in $\bbeta_{i,q}$ indicate
\emph{greater caries severity} among affected teeth.

Hence the induced probability mass function is
\[
P(C^{i}_{q}=0\mid \bx_i)=\pi^{(C)}_{iq},
\]
\[
P(C^{i}_{q}=u\mid \bx_i)
=
\{1-\pi^{(C)}_{iq}\}
\,P(C^{i}_{q}=u\mid C^{i}_{q}>0,\bx^{(s)}_i),
\qquad u=1,\ldots,C-1.
\]
Equivalently, for $u\ge 1$,
\[
P(C^{i}_{q}\le u\mid \bx_i)
=
\pi^{(C)}_{iq}
+
\{1-\pi^{(C)}_{iq}\}
P(C^{i}_{q}\le u\mid C^{i}_{q}>0,\bx^{(s)}_i).
\]
This decomposition makes clear that zeros arise only through the hurdle part,
whereas the proportional-odds model governs severity among positive outcomes.

We parameterize the cutpoints as
\[
\alpha_u=\sum_{j\le u}\exp(\delta_j)-\delta_0,
\qquad
u=1,\ldots,C-2,
\]
with $\delta_0,\delta_1,\ldots,\delta_{C-2}\in\mathbb R$, which guarantees
$\alpha_1<\cdots<\alpha_{C-2}$ automatically.

\paragraph*{Fluorosis model}
The fluorosis response $F^{i}_{q'}\in\{0,1,\ldots,F-1\}$ is modeled
analogously. The hurdle component models the probability of developing
fluorosis:
\[
P(F^{i}_{q'}>0\mid \bx^{(o)}_i)
=
\operatorname{logit}^{-1}\!\bigl((\brho_{i,q'})^{\top}\bx^{(o)}_i\bigr),
\]
so that $\pi^{(F)}_{iq'} = 1 - P(F^{i}_{q'}>0\mid \bx^{(o)}_i)$ is the
probability of remaining fluorosis-free. A positive coefficient in $\brho_{i,q'}$
indicates increased odds of developing fluorosis.
and for $v=1,\ldots,F-2$,
\[
P(F^{i}_{q'}\le v\mid F^{i}_{q'}>0,\bx^{(s)}_i)
=
\operatorname{logit}^{-1}\!\bigl(\gamma_v-(\btheta_{i,q'})^{\top}\bx^{(s)}_i\bigr).
\]
Then
\[
P(F^{i}_{q'}=0\mid \bx_i)=\pi^{(F)}_{iq'},
\]
\[
P(F^{i}_{q'}=v\mid \bx_i)
=
\{1-\pi^{(F)}_{iq'}\}
\,P(F^{i}_{q'}=v\mid F^{i}_{q'}>0,\bx^{(s)}_i),
\qquad v=1,\ldots,F-1,
\]
and for $v\ge 1$,
\[
P(F^{i}_{q'}\le v\mid \bx_i)
=
\pi^{(F)}_{iq'}
+
\{1-\pi^{(F)}_{iq'}\}
P(F^{i}_{q'}\le v\mid F^{i}_{q'}>0,\bx^{(s)}_i).
\]
The cutpoints are parameterized as
\[
\gamma_v=\sum_{j\le v}\exp(\xi_j)-\xi_0,
\qquad v=1,\ldots,F-2,
\]
with $\xi_0,\xi_1,\ldots,\xi_{F-2}\in\mathbb R$.

The proportional-odds assumption is made more flexible by allowing the linear
predictors to vary across individuals and locations \citep{Kang2023,sarkar2024analyzing}. We do not claim that this
eliminates all possible departures from proportional odds; rather, it softens
the usual restriction by allowing the effect of a predictor to vary over
subjects and spatial positions even though, conditional on a particular
location-specific coefficient vector, the positive-category model retains a
proportional-odds form. The hurdle component remains necessary even with these
location-specific coefficients because excess zeros and non-zero ordinal
severity address different aspects of the response distribution.

\subsection{\label{subsec:tucker} Linked Tucker tensor factorization}

Let $p_o$ and $p_s$ denote the numbers of predictors in the hurdle and severity
components, respectively. For notational simplicity below, write
\[
\bmu_{i,q}=\{\mu_{i,q,j}\}_{j=1}^{p_o},
\qquad
\brho_{i,q'}=\{\rho_{i,q',j}\}_{j=1}^{p_o},
\]
\[
\bbeta_{i,q}=\{\beta_{i,q,j}\}_{j=1}^{p_s},
\qquad
\btheta_{i,q'}=\{\theta_{i,q',j}\}_{j=1}^{p_s}.
\]
All four coefficient arrays are modeled through Tucker factorizations. In the
cross-sectional case, for multilinear ranks $(R_1^{(\ell)},R_2^{(\ell)},R_3^{(\ell)})$
with $\ell\in\{o,s\}$ denoting hurdle and severity, we write
\begin{align*}
\text{(Caries, Occurrence:)} \hspace{3em}
\mu_{i,q,j}
&=
\sum_{r_1=1}^{R_1^{(o)}}\sum_{r_2=1}^{R_2^{(o)}}\sum_{r_3=1}^{R_3^{(o)}}
 g^{(C,o)}_{r_1,r_2,r_3}
 a^{(1,o)}_{i,r_1}a^{(2,o)}_{q,r_2}a^{(3,o)}_{j,r_3},
\\
\text{(Fluorosis, Occurrence:)} \hspace{3em}
\rho_{i,q',j}
&=
\sum_{r_1=1}^{R_1^{(o)}}\sum_{r_2=1}^{\widetilde R_2^{(o)}}\sum_{r_3=1}^{R_3^{(o)}}
 g^{(F,o)}_{r_1,r_2,r_3}
 a^{(1,o)}_{i,r_1}b^{(2,o)}_{q',r_2}b^{(3,o)}_{j,r_3},
\\
\text{(Caries, Severity:)} \hspace{3em}
\beta_{i,q,j}
&=
\sum_{r_1=1}^{R_1^{(s)}}\sum_{r_2=1}^{R_2^{(s)}}\sum_{r_3=1}^{R_3^{(s)}}
 g^{(C,s)}_{r_1,r_2,r_3}
 a^{(1,s)}_{i,r_1}a^{(2,s)}_{q,r_2}a^{(3,s)}_{j,r_3},
\\
\text{(Fluorosis,  Severity:)} \hspace{3em}
\theta_{i,q',j}
&=
\sum_{r_1=1}^{R_1^{(s)}}\sum_{r_2=1}^{\widetilde R_2^{(s)}}\sum_{r_3=1}^{R_3^{(s)}}
 g^{(F,s)}_{r_1,r_2,r_3}
 a^{(1,s)}_{i,r_1}b^{(2,s)}_{q',r_2}b^{(3,s)}_{j,r_3}.
\end{align*}
This Tucker factorization decomposes each regression coefficient into contributions associated with the individual, spatial location (tooth-surface/zone), and predictor variable.
The shared subject factors $a^{(1,o)}$ and $a^{(1,s)}$ are the key linking
mechanism that makes this a \emph{joint} model for the two outcomes. 
They induce dependence between the caries and
fluorosis coefficient tensors by allowing the same latent subject profile to
influence both responses, with separate contributions for the occurrence and severity coefficients.
In contrast, we do not share the spatial factors
$a^{(2,\ell)}$ and $b^{(2,\ell)}$ because caries surfaces and fluorosis zones
are recorded
on different spatial resolutions. If both outcomes were measured on a
common grid, a shared spatial factor would be a natural extension. Likewise,
shared predictor factors could be imposed when scientific or design
considerations justify them.

This representation also clarifies the correlation structure. Two coefficient
entries associated with the same subject tend to be dependent because they
share the same latent subject scores in the first mode. Similarly, entries with
nearby locations or related predictors can be statistically coupled through the
shared spatial and predictor factors. The Tucker core tensors govern how these
mode-specific latent effects interact.

The multilinear ranks $(R_1,R_2,R_3)$ control the degree of compression. In
practice, they are treated as tuning parameters and can be chosen through a
small candidate grid together with predictive criteria or sensitivity analysis.
Because the present manuscript focuses on model formulation and application, we
leave the exact rank-selection workflow to the implementation details.

Under missing at random and parameter distinctness \citep{Little2019,Daniels2008}, inference uses the
observed-data likelihood formed only from observed tooth-location responses.
The low-rank structure is helpful in this setting because partially observed
subjects and spatial cells still inform one another through the shared latent
factors.

\section{\label{sec:longi} Longitudinal Extension with Time-Varying Predictors}

We now extend the paired hurdle-ordinal model to the longitudinal setting,
where the same subjects are examined at multiple ages $t=1,\ldots,T$ (in our
application, ages 9, 13, and 17 years) and covariates are updated at each
visit. Crucially, this is \emph{not} a collection of separate single-time-point
models: the data across all time points
enter a single joint model whose Tucker factorization
includes a time mode that is shared across visits. This allows the model to
borrow information across ages, subject to the low-rank structure imposed by
the time factor matrices. The number of assessable tooth surfaces (for caries)
and tooth zones (for fluorosis) may differ across ages as the dentition
develops; only observed tooth-location--time combinations contribute to the
likelihood.

\subsection{\label{subsec:longi_formulation} Longitudinal hurdle-ordinal formulation}

For caries, let $C^{i}_{q,t}\in\{0,1,\ldots,C-1\}$. The hurdle component at
visit $t$ models the probability of developing caries:
\[
P(C^{i}_{q,t}>0\mid \bx^{(o)}_{i,t})
=
\operatorname{logit}^{-1}\!\bigl((\bmu_{i,q,t})^{\top}\bx^{(o)}_{i,t}\bigr),
\]
where $\pi^{(C)}_{iqt} = 1 - P(C^{i}_{q,t}>0\mid \bx^{(o)}_{i,t})$ is the probability
of remaining caries-free at visit $t$.
and for $u=1,\ldots,C-2$,
\[
P(C^{i}_{q,t}\le u\mid C^{i}_{q,t}>0,\bx^{(s)}_{i,t})
=
\operatorname{logit}^{-1}\!\bigl(\alpha_u-(\bbeta_{i,q,t})^{\top}\bx^{(s)}_{i,t}\bigr).
\]
Thus,
\[
P(C^{i}_{q,t}=0\mid \bx_{i,t})=\pi^{(C)}_{iqt},
\]
\[
P(C^{i}_{q,t}=u\mid \bx_{i,t})
=
\{1-\pi^{(C)}_{iqt}\}
\,P(C^{i}_{q,t}=u\mid C^{i}_{q,t}>0,\bx^{(s)}_{i,t}),
\quad u=1,\ldots,C-1.
\]
An analogous specification is used for fluorosis:
\[
P(F^{i}_{q',t}>0\mid \bx^{(o)}_{i,t})
=
\operatorname{logit}^{-1}\!\bigl((\brho_{i,q',t})^{\top}\bx^{(o)}_{i,t}\bigr),
\]
where $\pi^{(F)}_{iq't} = 1 - P(F^{i}_{q',t}>0\mid \bx^{(o)}_{i,t})$.
\[
P(F^{i}_{q',t}\le v\mid F^{i}_{q',t}>0,\bx^{(s)}_{i,t})
=
\operatorname{logit}^{-1}\!\bigl(\gamma_v-(\btheta_{i,q',t})^{\top}\bx^{(s)}_{i,t}\bigr),
\quad v=1,\ldots,F-2.
\]

All four longitudinal coefficient arrays are assigned time-augmented Tucker
factorizations:
\begin{align*}
\text{(Caries, Occurrence:)}\qquad
\mu_{i,q,j,t}
&=
\sum_{r_1,r_2,r_3,r_4}
 g^{(C,o)}_{r_1,r_2,r_3,r_4}
 a^{(1,o)}_{i,r_1}a^{(2,o)}_{q,r_2}a^{(3,o)}_{j,r_3}a^{(4,o)}_{t,r_4},
\\
\text{(Fluorosis, Occurrence:)}\qquad
\rho_{i,q',j,t}
&=
\sum_{r_1,r_2,r_3,r_4}
 g^{(F,o)}_{r_1,r_2,r_3,r_4}
 a^{(1,o)}_{i,r_1}b^{(2,o)}_{q',r_2}b^{(3,o)}_{j,r_3}b^{(4,o)}_{t,r_4},
\\
\text{(Caries, Severity:)}\qquad
\beta_{i,q,j,t}
&=
\sum_{r_1,r_2,r_3,r_4}
 g^{(C,s)}_{r_1,r_2,r_3,r_4}
 a^{(1,s)}_{i,r_1}a^{(2,s)}_{q,r_2}a^{(3,s)}_{j,r_3}a^{(4,s)}_{t,r_4},
\\
\text{(Fluorosis, Severity:)}\qquad
\theta_{i,q',j,t}
&=
\sum_{r_1,r_2,r_3,r_4}
 g^{(F,s)}_{r_1,r_2,r_3,r_4}
 a^{(1,s)}_{i,r_1}b^{(2,s)}_{q',r_2}b^{(3,s)}_{j,r_3}b^{(4,s)}_{t,r_4}.
\end{align*}
The time factors $a^{(4,\ell)}$ and $b^{(4,\ell)}$ allow the coefficient
surfaces to evolve across visits while borrowing information across ages.
In plain terms, rather than fitting entirely independent models at each age,
the Tucker time mode links the age-9, age-13, and age-17 estimates through
shared latent structure, so that data from all three visits jointly inform
how covariate effects change over development.

\section{\label{sec:computation} Prior Specification and Computation}

In this section, we describe prior distributions for the factorization
components and outline posterior computation. Because the model contains
high-dimensional coefficient tensors, regularization is essential for stable
estimation.

\paragraph*{Core tensors}
For each outcome/component-specific core tensor element $g_k$, we use a
horseshoe prior \citep{carvalho2009handling}.:
\begin{align*}
 g_k\mid \lambda_k,\tau &\sim \mathcal N(0,\tau^2\lambda_k^2),
\\
 \lambda_k &\sim \mathrm{C}^+(0,1),
\\
 \tau &\sim \mathrm{C}^+(0,1).
\end{align*}
The local scale $\lambda_k$ allows element-specific shrinkage, while the global
scale $\tau$ controls overall sparsity. This is particularly useful because a
small subset of interactions in the Tucker core may dominate, while many others
are negligible.


\paragraph*{Mode matrices}
For the mode-matrix entries we use centered Gaussian priors,
\[
 a^{(m,\ell)}_{\cdot,r}\sim \mathcal N(0,\sigma_a^2),
 \qquad
 b^{(m,\ell)}_{\cdot,r}\sim \mathcal N(0,\sigma_b^2),
\]
with default choices $\sigma_a^2=\sigma_b^2=1$. These priors primarily serve
as scale regularizers for the latent factors. They do not by themselves fully
resolve the usual Tucker non-identifiability up to rotations and rescalings,
but they stabilize posterior exploration by discouraging extreme factor values.

\paragraph*{Cutpoints}
We place independent Gaussian priors on the unconstrained cutpoint parameters,
\[
\delta_j\sim \mathcal N(0,2^2),
\qquad
\xi_j\sim \mathcal N(0,2^2),
\]
which are weakly informative on the log-increment scale while preserving the
ordered-cutpoint constraints induced by the exponential-sum construction.

\paragraph*{Ranks} We pre-specify the ranks large enough and let the shrinkage prior allow data-driven automatic selection of the relevant ranks. We also maintain the rank $r_j$ to be less than the coefficient dimension for the $j$-th dimension.

\paragraph*{Occurrence and severity components}
The same Tucker-based prior structure is used for all four coefficient arrays:
$\bmu$ and $\brho$ in the occurrence (hurdle) component, and $\bbeta$ and $\btheta$ in the
severity component. 
The main distinction is scientific rather than
probabilistic: $\bmu$ and $\brho$ act on the odds of \emph{developing} disease
(positive values increase disease risk),
whereas $\bbeta$ and $\btheta$ act on severity among positive categories
(positive values indicate greater disease burden).

\paragraph*{Computation}
Posterior computation is carried out using the No-U-Turn Sampler (NUTS)
\citep{hoffman2014no}, implemented in NumPyro \citep{bingham2019pyro}.
Automatic differentiation is especially convenient here because the log
posterior involves many structured tensor operations. All posterior
computations are run on GPU-enabled hardware through JAX/NumPyro.
In our empirical analyses, we discard the first 5{,}000 iterations as burn-in and
retain the next 5{,}000 posterior draws for inference. 


\section{\label{sec:inference} Inference}

The fitted model yields subject- and location-specific posterior draws for the
coefficient vectors, which are scientifically rich but difficult to summarize
directly. Our primary inferential target is therefore a projected,
population-level coefficient summary for each spatial unit. The idea is to take
the posterior draw of the individualized linear predictor across subjects and
project it back onto the design space. This produces the best linear
least-squares approximation, in the observed covariate space, to the collection
of subject-specific effects at that location.

Formally, for each posterior draw $m$ and caries location $q\in\mathcal Q_C$,
let
\[
\bc^{(m)}_q
=
\bigl((\bmu^{(m)}_{1,q})^{\top}\bx_1,\ldots,(\bmu^{(m)}_{n,q})^{\top}\bx_n\bigr)^{\top}
\]
be the vector of linear predictors for the caries occurrence outcome at
location $q$ across all individuals in posterior draw $m$. If $\bX_o$ denotes
the occurrence-model design matrix, including its intercept column, define
\[
\bP_o=(\bX_o^{\top}\bX_o)^{-1}\bX_o^{\top},
\qquad
\widetilde{\bmu}^{(m)}_q=\bP_o\bc_q^{(m)}.
\]
Thus, $\widetilde{\bmu}^{(m)}_q$ is the projected coefficient vector at
location $q$. Analogous constructions are used for fluorosis and for the
severity coefficients of both outcomes. For a severity model, the projection
uses its design matrix $\bX_s$ without an intercept, consistently with the
proportional-odds specification.

This projection step is preferable to directly averaging the raw tensor
coefficients because the coefficient tensors act through the linear predictors,
not in isolation.
The projection therefore summarizes the covariate effect on
the scale at which the model is identified and interpreted. It also yields a
distinct projected coefficient vector for each spatial unit, which is then
amenable to scientifically meaningful aggregation.

We use two levels of summarization. First, within a given tooth we average over
its constituent surfaces (for caries) or zones (for fluorosis) to obtain
tooth-level posterior summaries. Second, we aggregate across teeth within a
common anatomical class (incisor, canine, premolar, molar) using
\emph{Wasserstein barycenters} of the corresponding posterior distributions.
The Wasserstein barycenter is a generalization of the arithmetic mean to the
space of probability distributions \citep{agueh2011barycenters}: it finds the distribution that minimizes
the average squared Wasserstein (optimal transport) distance to a collection of
input distributions. In our context, each tooth-type class contributes a
posterior distribution over the projected coefficient vector, and the barycenter
aggregates these distributions in a geometrically meaningful way that respects
the shape and spread of each contributing distribution rather than simply
averaging point estimates. We likewise form surface-level barycenters for
caries and zone-level barycenters for fluorosis by aggregating over teeth while
preserving the anatomical index. More generally, for a selected set of
locations $\widetilde{\mathcal Q}\subset\mathcal Q$, let
$\widetilde{\bmu}_{\widetilde{\mathcal Q}}$ denote the Wasserstein barycenter
of the empirical posterior distributions of $\widetilde{\bmu}_q$, for
$q\in\widetilde{\mathcal Q}$. No parametric distribution is imposed in this
aggregation; the barycenter is computed directly from the posterior draws applying the R package {\tt waspr} \citep{cremers2020waspr}.

Throughout the tables, the reported intervals are equal-tail 95\% posterior
credible intervals obtained from the empirical quantiles of the corresponding
projected or barycentered posterior draws. They quantify posterior uncertainty
for the aggregated effect after the relevant averaging and barycenter
operations. Thus, when locations are aggregated, the intervals describe the
posterior distribution of $\widetilde{\bmu}_{\widetilde{\mathcal Q}}$; they do
not represent cross-subject variability in the individualized coefficients.

\section{\label{sec:application} Dental Data Application}

We demonstrate our proposed methodology with two analyses using the IFS data.
The first is a joint
cross-sectional analysis of caries at age 5 and fluorosis at age 9, both
regressed on covariates measured from birth through age 5. The two outcomes
are analyzed together in a single fitted model via the linked Tucker
factorization described in Section~\ref{sec:model}: the shared subject-mode factors couple
the caries and fluorosis coefficient tensors within a single joint likelihood,
and all model parameters are estimated simultaneously. The caries and fluorosis
results are presented in separate tables (Tables~\ref{tab:C5} and \ref{tab:F9})
because they operate on distinct spatial grids (tooth surfaces vs.\ tooth
zones), but they arise from a single joint model fit. The second application
is a joint longitudinal analysis of caries and fluorosis at ages 9, 13, and
17, regressed on time-varying covariates.

Here, the age-5/age-9 pairing reflects the distinct developmental windows of the two
conditions. Caries at age 5 captures decay occurring on the primary (baby)
dentition, when enamel formation is complete and carious lesions reflect
cumulative dietary and behavioral exposures during early childhood. Fluorosis
at age 9 captures enamel changes on the permanent incisors and first molars,
which mineralize during the first several years of life and are the teeth most
sensitive to fluoride intake during that window. Regressing both outcomes on
covariates measured from birth to age 5 allows the model to capture the common
early-life exposure period that shapes both conditions.

\paragraph*{Interpretation of the Occurrence and Severity Components}
Note that our ordinal hurdle model decomposes each outcome into:
(i) an \emph{occurrence component} (hurdle) that models the odds of
\emph{developing} disease (i.e., crossing from disease-absent to
disease-present), and
(ii) a \emph{severity component} that models the ordinal disease level
conditional on being non-zero.

Therefore, in the tables below, {positive effects in the Occurrence
block indicate higher odds of developing disease (greater disease risk)},
whereas {negative effects indicate lower odds of developing disease
(protective)}. In the Severity block, the sign is interpreted through the
proportional-odds parameterization: {positive effects shift mass toward
higher non-zero disease categories (greater severity), while negative effects
indicate lower severity among those affected}.

Throughout, we interpret an effect as meaningful when the reported 95\%
posterior credible interval excludes zero (shown in bold in the tables). This
allows age-specific occurrence/severity inference, where separating the two
components improves interpretability under heavy zero inflation. Because the
two components are modeled jointly and results are presented for both, the
tables for each analysis show the Occurrence block first, followed by the
Severity block.

\subsection{\label{subsec:age5age9} Age-5 Caries and Age-9 Fluorosis Regressed on Baseline Age-5 Covariates}

We run a joint analysis of age-5 caries (primary dentition) and age-9
fluorosis (permanent dentition) with all predictors reflecting the period from
birth to age 5. Because the teeth present at age 5 are baby teeth,
tooth-type-specific analyses are not performed for this cross-sectional
application.

\begin{figure}[htbp]
    \centering
    \includegraphics[width=0.8\textwidth, trim=0cm 0cm 0cm 2cm, clip=true]{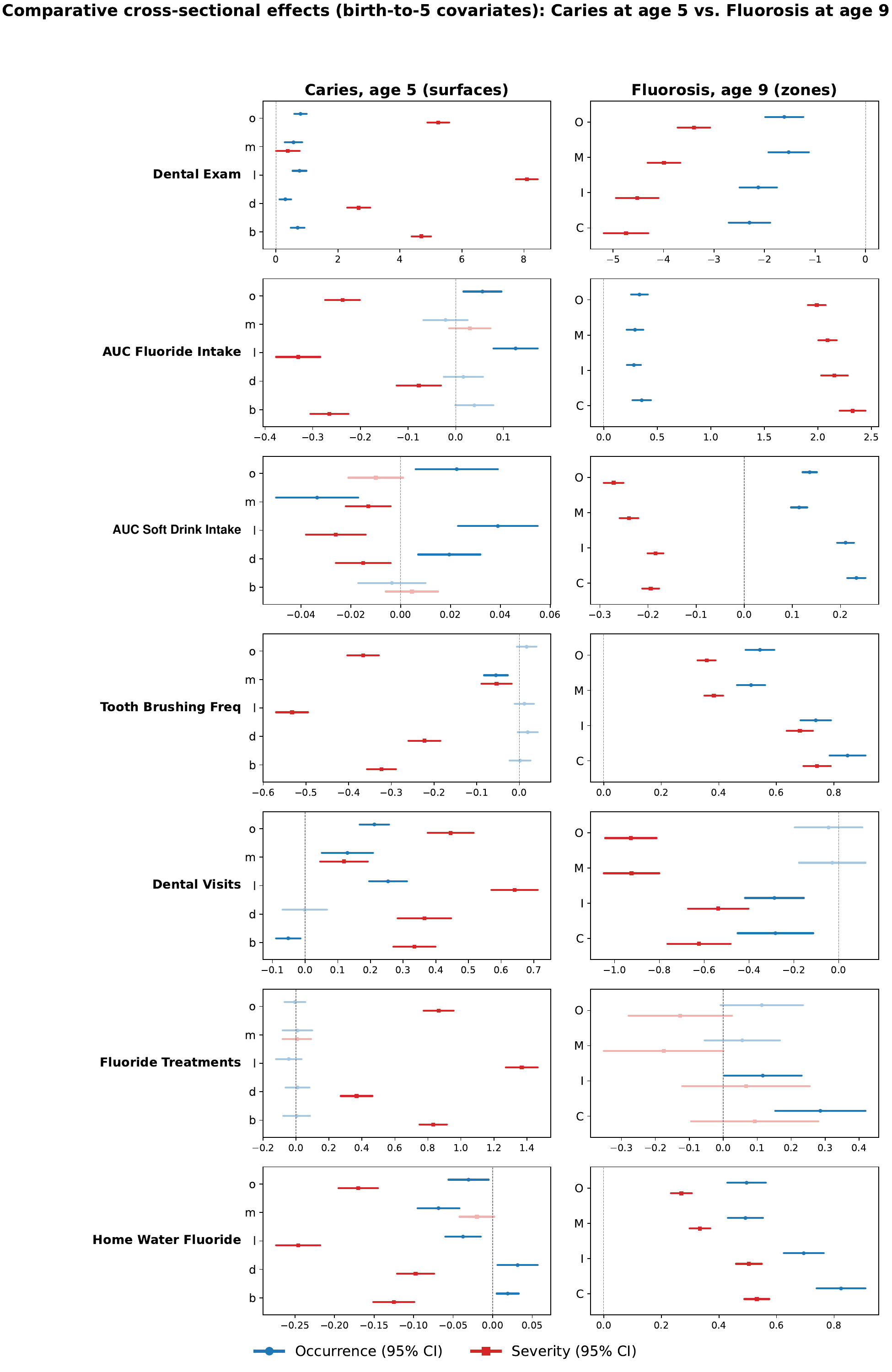}
    \caption{Comparative cross-sectional effects (birth-to-5 covariates): Caries at age 5 (surfaces, left panel) vs.\ Fluorosis at age 9 (zones, right panel), shown for all 7 predictors. Occurrence (blue circle) and Severity (red square) markers show the posterior mean with 95\% credible interval; the darker/solid marker indicates the interval excludes zero. Summarizes Tables~\ref{tab:C5} and \ref{tab:F9}.}
    \label{fig:comparative_age5_age9}
\end{figure}

Table~\ref{tab:C5} (caries, tooth surfaces) and Table~\ref{tab:F9} (fluorosis, tooth zones)
together summarize the single joint model
and are further illustrated in Figure~\ref{fig:comparative_age5_age9}.
In each table, the Occurrence block
(upper rows) is presented first, followed by the Severity block (lower rows),
consistent with the logical sequence of disease development. All predictor
variables reflect cumulative exposure from birth to age 5; AUC quantities
represent area-under-the-curve summaries of daily exposure integrated over this
period (see Section~\ref{sec:variables} for variable definitions).

\begin{table}[ht]
\centering
\caption{Joint model results: Wasserstein barycenter-based summarization for different tooth surfaces (Caries, age 5). This table is part of a single joint model also including fluorosis at age 9 (Table~\ref{tab:F9}). The Occurrence block (upper rows) shows odds of developing caries; the Severity block (lower rows) shows severity among affected teeth. All predictors reflect cumulative exposure from birth to age 5 (AUC = area under the curve). Intervals are 95\% posterior credible intervals; those excluding zero are in bold.}
\scriptsize
\setlength{\tabcolsep}{3pt}
\renewcommand{\arraystretch}{0.8}
\resizebox{\textwidth}{!}{
\begin{tabular}{lrrrrrrr}
\toprule
 & \shortstack{Dental\\Exam}   & \shortstack{AUC Fluoride\\Intake (mg),\\birth--5 yrs}   & \shortstack{AUC Soft Drink\\Intake (oz),\\birth--5 yrs}   & \shortstack{Tooth Brushing\\Freq (per day),\\avg birth--5 yrs}   & \shortstack{Dental Visits\\(past 6 mo),\\avg birth--5 yrs}   & \shortstack{Fluoride\\Treatments\\(6 mo avg)}   & \shortstack{Home Water\\Fluoride (ppm),\\avg birth--5 yrs} \\ \\
\midrule
\multicolumn{8}{l}{\textit{Occurrence component: odds of developing caries}} \\[2pt]
  Caries 5yr Occurrence b & \textbf{(0.479, 0.923)} & (-0.001, 0.079) & (-0.017, 0.010) & (-0.023, 0.026) & \textbf{(-0.089, -0.014)} & (-0.077, 0.085) & \textbf{(0.005, 0.033)} \\
  Caries 5yr Occurrence d & \textbf{(0.116, 0.492)} & (-0.025, 0.057) & \textbf{(0.007, 0.032)} & (-0.004, 0.043) & (-0.069, 0.067) & (-0.063, 0.082) & \textbf{(0.006, 0.057)} \\
  Caries 5yr Occurrence l & \textbf{(0.537, 0.987)} & \textbf{(0.079, 0.172)} & \textbf{(0.023, 0.055)} & (-0.011, 0.034) & \textbf{(0.196, 0.312)} & (-0.121, 0.034) & \textbf{(-0.060, -0.015)} \\
  Caries 5yr Occurrence m & \textbf{(0.285, 0.849)} & (-0.068, 0.025) & \textbf{(-0.050, -0.017)} & \textbf{(-0.083, -0.027)} & \textbf{(0.051, 0.208)} & (-0.081, 0.099) & \textbf{(-0.095, -0.042)} \\
  Caries 5yr Occurrence o & \textbf{(0.593, 0.994)} & \textbf{(0.016, 0.096)} & \textbf{(0.006, 0.039)} & (-0.006, 0.040) & \textbf{(0.167, 0.257)} & (-0.069, 0.057) & \textbf{(-0.056, -0.005)} \\ 
\addlinespace
\multicolumn{8}{l}{\textit{Severity component: disease severity among caries-affected teeth}} \\[2pt]
Caries 5yr Severity b & \textbf{(4.379, 5.004)} & \textbf{(-0.305, -0.225)} & (-0.006, 0.015) & \textbf{(-0.357, -0.289)} & \textbf{(0.270, 0.399)} & \textbf{(0.749, 0.915)} & \textbf{(-0.151, -0.099)} \\
Caries 5yr Severity d & \textbf{(2.300, 3.044)} & \textbf{(-0.124, -0.031)} & \textbf{(-0.026, -0.004)} & \textbf{(-0.260, -0.185)} & \textbf{(0.283, 0.447)} & \textbf{(0.271, 0.464)} & \textbf{(-0.121, -0.074)} \\
Caries 5yr Severity l & \textbf{(7.741, 8.445)} & \textbf{(-0.377, -0.284)} & \textbf{(-0.038, -0.014)} & \textbf{(-0.570, -0.495)} & \textbf{(0.570, 0.712)} & \textbf{(1.272, 1.466)} & \textbf{(-0.274, -0.218)} \\
Caries 5yr Severity m & \textbf{(0.003, 0.765)} & (-0.014, 0.073) & \textbf{(-0.022, -0.004)} & \textbf{(-0.089, -0.018)} & \textbf{(0.046, 0.192)} & (-0.081, 0.092) & (-0.042, 0.002) \\
Caries 5yr Severity o & \textbf{(4.885, 5.590)} & \textbf{(-0.274, -0.201)} & (-0.021, 0.001) & \textbf{(-0.403, -0.329)} & \textbf{(0.375, 0.516)} & \textbf{(0.774, 0.956)} & \textbf{(-0.195, -0.145)} \\ 
 \bottomrule
\end{tabular}}
\label{tab:C5}
\end{table}

\begin{table}[ht]
\centering
\caption{Joint model results: Wasserstein barycenter-based summarization for different tooth zones (Fluorosis, age 9). This table is part of a single joint model also including caries at age 5 (Table~\ref{tab:C5}). The Occurrence block (upper rows) shows odds of developing fluorosis; the Severity block (lower rows) shows severity among affected teeth. All predictors reflect cumulative exposure from birth to age 5 (AUC = area under the curve). Tooth zones: C = cervical, I = incisal, M = middle, O = occlusal. Intervals are 95\% posterior credible intervals; those excluding zero are in bold.}
\scriptsize
\setlength{\tabcolsep}{3pt}
\renewcommand{\arraystretch}{0.8}
\resizebox{\textwidth}{!}{
\begin{tabular}{lrrrrrrr}
\toprule
 & \shortstack{Dental\\Exam}   & \shortstack{AUC Fluoride\\Intake (mg),\\birth--5 yrs}   & \shortstack{AUC Soft Drink\\Intake (oz),\\birth--5 yrs}   & \shortstack{Tooth Brushing\\Freq (per day),\\avg birth--5 yrs}   & \shortstack{Dental Visits\\(past 6 mo),\\avg birth--5 yrs}   & \shortstack{Fluoride\\Treatments\\(6 mo avg)}   & \shortstack{Home Water\\Fluoride (ppm),\\avg birth--5 yrs} \\ \\
\midrule
\multicolumn{8}{l}{\textit{Occurrence component: odds of developing fluorosis}} \\[2pt]
  Fluorosis 9yr Occurrence C & \textbf{(-2.708, -1.887)} & \textbf{(0.268, 0.441)} & \textbf{(0.214, 0.252)} & \textbf{(0.785, 0.910)} & \textbf{(-0.451, -0.113)} & \textbf{(0.153, 0.419)} & \textbf{(0.739, 0.909)} \\
  Fluorosis 9yr Occurrence I & \textbf{(-2.494, -1.753)} & \textbf{(0.216, 0.347)} & \textbf{(0.193, 0.228)} & \textbf{(0.684, 0.790)} & \textbf{(-0.418, -0.155)} & \textbf{(0.002, 0.231)} & \textbf{(0.625, 0.764)} \\
  Fluorosis 9yr Occurrence M & \textbf{(-1.921, -1.121)} & \textbf{(0.214, 0.370)} & \textbf{(0.097, 0.131)} & \textbf{(0.463, 0.561)} & (-0.177, 0.120) & (-0.055, 0.167) & \textbf{(0.431, 0.553)} \\
  Fluorosis 9yr Occurrence O & \textbf{(-1.986, -1.231)} & \textbf{(0.255, 0.412)} & \textbf{(0.121, 0.151)} & \textbf{(0.493, 0.593)} & (-0.196, 0.106) & (-0.008, 0.235) & \textbf{(0.429, 0.563)} \\ 
\addlinespace
\multicolumn{8}{l}{\textit{Severity component: disease severity among fluorosis-affected teeth}} \\[2pt]
Fluorosis 9yr Severity C & \textbf{(-5.183, -4.301)} & \textbf{(2.203, 2.445)} & \textbf{(-0.212, -0.177)} & \textbf{(0.694, 0.789)} & \textbf{(-0.764, -0.482)} & (-0.095, 0.280) & \textbf{(0.488, 0.575)} \\
Fluorosis 9yr Severity I & \textbf{(-4.942, -4.098)} & \textbf{(2.031, 2.279)} & \textbf{(-0.201, -0.168)} & \textbf{(0.636, 0.727)} & \textbf{(-0.672, -0.402)} & (-0.121, 0.255) & \textbf{(0.459, 0.549)} \\
Fluorosis 9yr Severity M & \textbf{(-4.314, -3.667)} & \textbf{(2.004, 2.176)} & \textbf{(-0.259, -0.220)} & \textbf{(0.350, 0.415)} & \textbf{(-1.048, -0.800)} & (-0.352, 0.001) & \textbf{(0.298, 0.370)} \\
Fluorosis 9yr Severity O & \textbf{(-3.721, -3.076)} & \textbf{(1.907, 2.072)} & \textbf{(-0.292, -0.251)} & \textbf{(0.326, 0.389)} & \textbf{(-1.042, -0.812)} & (-0.279, 0.026) & \textbf{(0.233, 0.306)} \\ 
 \bottomrule
\end{tabular}}
\label{tab:F9}
\end{table}

\paragraph*{Caries at age 5 (surface-level barycenters): Occurrence component}
Table~\ref{tab:C5} reveals distinct patterns for caries. In the
\emph{Occurrence} block, dental exam status shows strong positive occurrence
effects across all surfaces, indicating higher odds of developing caries among
children with a recorded non-study dental examination; this most likely reflects
reactive care-seeking in children already at elevated caries risk. Dental
visits show positive occurrence effects on the lingual, mesial, and occlusal
surfaces for the same reason. Soft drink intake shows positive occurrence effects on
the distal and lingual surfaces, consistent with sugary beverage consumption
as a caries risk factor \citep{Marshall2003,Levy2003}. Home water fluoride shows
negative occurrence effects on the buccal, lingual, and occlusal surfaces,
consistent with the protective role of community water fluoridation. Fluoride
treatment shows negative effects for buccal occurrence, again consistent with
a protective role when administered preventively.

\paragraph*{Caries at age 5 (surface-level barycenters): Severity component}
In the \emph{Severity} block of Table~\ref{tab:C5}, effects are conditional on
the subset of teeth with observed caries at age 5; the proportional-odds
cutpoints, rather than a separate regression intercept, capture baseline
severity.
Fluoride intake AUC shows negative severity effects on several surfaces
(buccal, distal, lingual, occlusal), indicating lower caries severity among
affected teeth in children with higher cumulative fluoride intake---consistent
with the well-established protective role of fluoride on enamel
demineralization \citep{Levy2003,mukherjee2024modeling}. Similarly, home water fluoride shows negative
severity effects across most surfaces. Fluoride treatment shows positive
severity effects on buccal, distal, lingual, and occlusal surfaces, again
likely reflecting reactive administration to children with more severe
underlying disease. Dental visits show positive severity effects, consistent
with visits being associated with more severe disease already present. Tooth
brushing frequency shows negative severity effects across surfaces, indicating
lower severity among children with better oral hygiene.

\paragraph*{Fluorosis at age 9 (zone-level barycenters): Occurrence component}
Table~\ref{tab:F9} shows strong and highly consistent signals across all four
tooth zones (C/I/M/O). In the \emph{Occurrence} block, fluoride intake AUC
(\texttt{AUC Fluoride Intake}) and home water fluoride
(\texttt{Home Water Fluoride}) show positive non-zero effects across all zones,
indicating increased odds of developing fluorosis with higher early-life
fluoride exposure---consistent with the well-established role of systemic
fluoride during enamel formation as the primary determinant of fluorosis risk
\citep{Levy2003,Kang2023,sarkar2024analyzing}. Soft drink intake AUC and tooth-brushing frequency also
show positive non-zero occurrence effects, indicating increased odds of
developing fluorosis in association with these behavioral factors. Dental exam
status shows negative occurrence effects across all zones, suggesting that
children with recorded non-study dental examinations had lower odds of developing
fluorosis, possibly reflecting receipt of fluoride-exposure counseling or
preventive guidance \citep{Broffitt2013}.

\paragraph*{Fluorosis at age 9 (zone-level barycenters): Severity component}
In the \emph{Severity} block, the large negative intercepts reflect the
overall rarity of high-severity fluorosis. Fluoride intake AUC shows
consistently positive non-zero effects across all zones, indicating greater
fluorosis severity among affected teeth in children with higher cumulative
fluoride intake. Home water fluoride is likewise positive and significant across
zones. Dental visits show negative severity effects (lower severity in children
with more dental care). Soft drink intake shows consistently negative severity
effects. Thus, greater soft drink intake is associated with higher fluorosis
occurrence but lower severity among affected teeth. One possible explanation is
that soft drinks displace fluoridated water or other fluoride-containing
beverages, although this observational association should not be interpreted
causally \citep{Marshall2003}.

Thus, fluoride intake and home water fluoride are associated with \emph{both}
higher odds of developing fluorosis (positive Occurrence) \emph{and} greater
disease severity (positive Severity), producing consistent risk signals in
both model components. Dental care indicators are protective in the Occurrence
block and associated with lower severity, suggesting a coherent preventive
role.

\paragraph*{Summary of the cross-sectional joint analysis}
The age-5/age-9 joint analysis demonstrates that associations differ not only
between caries and fluorosis, but also between occurrence and severity
components. For fluorosis, higher fluoride exposure is associated with
\emph{increased} odds of developing fluorosis (positive Occurrence) \emph{and}
greater severity among affected teeth (positive Severity)---a consistent
risk pattern in both components. For caries, fluoride exposure is associated
with \emph{lower} caries severity (negative Severity) and lower occurrence
risk for water fluoride (negative Occurrence), while soft drink intake increases
caries occurrence risk. Although some predictors have effects in the same direction for occurrence and severity, others are associated with only one component or differ substantially in magnitude across components. Separating disease occurrence from disease severity is therefore necessary to identify whether a predictor is associated with developing disease, its severity among affected teeth, or both. The joint modeling framework allows these patterns to be examined simultaneously
while borrowing strength across spatial units
\citep{sarkar2024analyzing,mukherjee2024modeling}.


\subsection{\label{subsec:longitudinal_app} Longitudinal Caries and Fluorosis at Ages 9, 13, and 17 Regressed on Time-Varying Covariates}

The longitudinal analysis fits a single joint model covering examination ages 9, 13, and 17 simultaneously for both caries and fluorosis (see Section~\ref{sec:longi}). This is \emph{not} a set of separate models per age: the Tucker factorization includes a shared time mode so that the age-9, age-13, and age-17 coefficient surfaces are all estimated jointly, borrowing information across developmental stages. Predictors are time-varying---for each examination age, the predictor values reflect cumulative exposures during the period \emph{preceding} that exam (e.g., the age-9 exam uses exposures from approximately ages 5--9, while the age-13 exam uses exposures from approximately ages 9--13). The number of assessable tooth surfaces and zones increases with age as the permanent dentition erupts; only tooth--surface or tooth--zone combinations observed at a given exam contribute to the likelihood at that time point (see Section~\ref{sec:variables}). Tooth-type-specific analyses (incisor, canine, premolar, molar) are additionally presented for the longitudinal analysis.

\begin{table}[ht]
\centering
\caption{Longitudinal joint model: Wasserstein barycenter-based summarization for different tooth zones (Fluorosis) and examination ages (9, 13, 17 years). Occurrence block (upper rows): odds of developing fluorosis. Severity block (lower rows): disease severity among fluorosis-affected teeth. Tooth zones: C = cervical, I = incisal, M = middle, O = occlusal. Predictors are time-varying, measured in the period preceding each examination. AUC = area under the curve for the relevant period. Intervals are 95\% posterior credible intervals; those excluding zero are in bold.}
\scriptsize
\setlength{\tabcolsep}{3pt}
\renewcommand{\arraystretch}{0.8}
\resizebox{\textwidth}{!}{
\begin{tabular}{lrrrrrrr}
\toprule
 & \shortstack{Dental\\Exam}   & \shortstack{AUC Fluoride\\Intake (mg)}   & \shortstack{AUC Soft Drink\\Intake (oz)}   & \shortstack{Tooth Brushing\\Freq/Day Avg}   & \shortstack{Dental Visits\\Past 6 mo Avg}   & \shortstack{Fluoride\\Treatment 6 mo Avg}   & \shortstack{Home Water\\Fluoride (ppm) Avg} \\ \\
\midrule
\multicolumn{8}{l}{\textit{Occurrence component: odds of developing fluorosis}} \\[2pt]
  Fluorosis 9yr Occurrence C & (-0.042, 0.000) & (-0.018, 0.008) & \textbf{(0.000, 0.004)} & (-0.001, 0.009) & (-0.018, 0.007) & (-0.010, 0.012) & \textbf{(0.008, 0.023)} \\ 
  Fluorosis 9yr Occurrence I & (-0.021, 0.024) & (-0.015, 0.013) & \textbf{(0.002, 0.006)} & (-0.005, 0.006) & (-0.013, 0.014) & (-0.018, 0.013) & \textbf{(0.011, 0.028)} \\ 
  Fluorosis 9yr Occurrence M & (-0.007, 0.034) & \textbf{(0.005, 0.030)} & \textbf{(0.001, 0.004)} & (-0.008, 0.001) & (-0.012, 0.011) & (-0.011, 0.010) & (-0.010, 0.005) \\ 
  Fluorosis 9yr Occurrence O & (-0.025, 0.015) & (-0.007, 0.019) & (-0.002, 0.001) & (-0.008, 0.002) & (-0.018, 0.009) & (-0.009, 0.016) & (-0.012, 0.005) \\ 
  Fluorosis 13yr Occurrence C & (-0.048, 0.047) & \textbf{(0.013, 0.034)} & (-0.003, 0.001) & (-0.002, 0.009) & \textbf{(0.000, 0.040)} & \textbf{(-0.046, -0.007)} & (-0.016, 0.005) \\ 
  Fluorosis 13yr Occurrence I & \textbf{(0.050, 0.155)} & \textbf{(0.006, 0.027)} & \textbf{(0.000, 0.004)} & (-0.002, 0.011) & \textbf{(0.028, 0.073)} & \textbf{(-0.063, -0.018)} & (-0.022, 0.002) \\ 
  Fluorosis 13yr Occurrence M & \textbf{(0.025, 0.125)} & (-0.008, 0.012) & (-0.002, 0.002) & (-0.005, 0.007) & \textbf{(0.008, 0.043)} & (-0.034, 0.000) & (-0.017, 0.000) \\ 
  Fluorosis 13yr Occurrence O & (-0.008, 0.098) & \textbf{(0.005, 0.027)} & (-0.001, 0.002) & (-0.003, 0.007) & (-0.006, 0.028) & (-0.039, 0.002) & \textbf{(-0.024, -0.003)} \\ 
  Fluorosis 17yr Occurrence C & (-0.069, 0.041) & \textbf{(0.008, 0.022)} & (-0.002, 0.002) & \textbf{(0.005, 0.015)} & (-0.009, 0.015) & (-0.009, 0.007) & \textbf{(-0.014, -0.007)} \\ 
  Fluorosis 17yr Occurrence I & (-0.112, -0.023) & \textbf{(0.009, 0.024)} & (-0.003, 0.002) & \textbf{(0.005, 0.015)} & (-0.020, 0.005) & (-0.003, 0.014) & (-0.003, 0.016) \\ 
  Fluorosis 17yr Occurrence M & (-0.044, 0.061) & (-0.003, 0.011) & (-0.003, 0.001) & (-0.011, 0.002) & (-0.010, 0.013) & (-0.002, 0.012) & (-0.018, 0.001) \\ 
  Fluorosis 17yr Occurrence O & (-0.051, 0.064) & (-0.009, 0.005) & (-0.004, 0.001) & (-0.007, 0.003) & (-0.016, 0.008) & (-0.008, 0.009) & (-0.002, 0.019) \\ 
\addlinespace
\multicolumn{8}{l}{\textit{Severity component: disease severity among fluorosis-affected teeth}} \\[2pt]
Fluorosis 9yr Severity C & \textbf{(0.012, 0.042)} & \textbf{(0.002, 0.009)} & (-0.001, 0.001) & (-0.004, 0.001) & \textbf{(0.002, 0.010)} & (-0.001, 0.004) & (-0.002, 0.004) \\ 
  Fluorosis 9yr Severity I & (-0.012, 0.022) & \textbf{(-0.009, -0.001)} & \textbf{(0.000, 0.002)} & \textbf{(0.000, 0.004)} & \textbf{(0.001, 0.009)} & (-0.001, 0.005) & (-0.004, 0.003) \\ 
  Fluorosis 9yr Severity M & \textbf{(0.002, 0.033)} & \textbf{(-0.012, -0.004)} & \textbf{(0.001, 0.003)} & \textbf{(0.001, 0.005)} & \textbf{(0.006, 0.012)} & (-0.002, 0.004) & \textbf{(0.000, 0.007)} \\ 
  Fluorosis 9yr Severity O & (-0.013, 0.030) & (-0.004, 0.005) & (-0.001, 0.001) & (-0.000, 0.004) & \textbf{(0.001, 0.008)} & \textbf{(0.001, 0.007)} & (-0.001, 0.006) \\ 
  Fluorosis 13yr Severity C & (-0.049, 0.017) & \textbf{(-0.019, -0.003)} & (-0.001, 0.001) & (-0.002, 0.005) & (-0.006, 0.007) & (-0.009, 0.006) & (-0.001, 0.015) \\ 
  Fluorosis 13yr Severity I & \textbf{(0.005, 0.070)} & (-0.001, 0.013) & (-0.001, 0.001) & (-0.001, 0.006) & (-0.003, 0.010) & (-0.008, 0.011) & (-0.008, 0.010) \\ 
  Fluorosis 13yr Severity M & \textbf{(0.014, 0.074)} & (-0.007, 0.005) & \textbf{(-0.002, -0.000)} & (-0.005, 0.002) & (-0.012, 0.000) & \textbf{(-0.028, -0.014)} & (-0.013, 0.005) \\ 
  Fluorosis 13yr Severity O & (-0.011, 0.060) & (-0.009, 0.007) & (-0.001, 0.001) & (-0.005, 0.004) & \textbf{(0.001, 0.014)} & (-0.015, 0.001) & (-0.012, 0.009) \\ 
  Fluorosis 17yr Severity C & (-0.023, 0.030) & (-0.002, 0.009) & \textbf{(-0.001, -0.000)} & (-0.010, 0.000) & \textbf{(0.008, 0.021)} & \textbf{(-0.020, -0.004)} & \textbf{(-0.019, -0.004)} \\ 
  Fluorosis 17yr Severity I & (-0.040, 0.018) & (-0.009, 0.002) & (-0.001, 0.000) & (-0.006, 0.004) & (-0.007, 0.008) & (-0.007, 0.009) & (-0.004, 0.011) \\ 
  Fluorosis 17yr Severity M & \textbf{(-0.069, -0.018)} & \textbf{(-0.011, -0.001)} & (-0.001, 0.000) & (-0.001, 0.007) & (-0.008, 0.008) & \textbf{(0.012, 0.027)} & (-0.006, 0.010) \\ 
  Fluorosis 17yr Severity O & \textbf{(0.004, 0.058)} & (-0.002, 0.010) & \textbf{(-0.002, -0.001)} & \textbf{(-0.011, -0.001)} & (-0.007, 0.009) & (-0.009, 0.008) & \textbf{(-0.015, -0.003)} \\ 
 \bottomrule
\end{tabular}}
\label{tab:F17long}
\end{table}

\begin{table}[ht]
\centering
\caption{Longitudinal joint model: Wasserstein barycenter-based summarization for different tooth surfaces (Caries) and examination ages (9, 13, 17 years). Occurrence block (upper rows): odds of developing caries. Severity block (lower rows): disease severity among caries-affected teeth. Tooth surfaces: b = buccal, d = distal, l = lingual, m = mesial, o = occlusal, od = occluso-distal, om = occluso-mesial. Predictors are time-varying, measured in the period preceding each examination. AUC = area under the curve for the relevant period. Intervals are 95\% posterior credible intervals; those excluding zero are in bold.}
\scriptsize
\setlength{\tabcolsep}{3pt}
\renewcommand{\arraystretch}{0.8}
\resizebox{\textwidth}{!}{
\begin{tabular}{lrrrrrrr}
\toprule
 & \shortstack{Dental\\Exam}   & \shortstack{AUC Fluoride\\Intake (mg)}   & \shortstack{AUC Soft Drink\\Intake (oz)}   & \shortstack{Tooth Brushing\\Freq/Day Avg}   & \shortstack{Dental Visits\\Past 6 mo Avg}   & \shortstack{Fluoride\\Treatment 6 mo Avg}   & \shortstack{Home Water\\Fluoride (ppm) Avg} \\ \\
\midrule
\multicolumn{8}{l}{\textit{Occurrence component: odds of developing caries}} \\[2pt]
  Caries 9yr Occurrence b & \textbf{(-0.077, -0.012)} & \textbf{(-0.022, -0.002)} & \textbf{(-0.004, -0.001)} & (-0.006, 0.003) & \textbf{(0.006, 0.037)} & (-0.027, 0.001) & \textbf{(0.002, 0.017)} \\ 
  Caries 9yr Occurrence d & \textbf{(-0.061, -0.002)} & (-0.006, 0.014) & \textbf{(-0.004, -0.000)} & (-0.002, 0.006) & (-0.002, 0.030) & (-0.017, 0.010) & (-0.005, 0.010) \\ 
  Caries 9yr Occurrence l & \textbf{(0.021, 0.091)} & (-0.006, 0.017) & (-0.002, 0.002) & \textbf{(-0.011, -0.000)} & (-0.021, 0.011) & (-0.015, 0.014) & (-0.006, 0.011) \\ 
  Caries 9yr Occurrence m & \textbf{(0.031, 0.086)} & \textbf{(0.020, 0.041)} & (-0.001, 0.002) & \textbf{(-0.014, -0.005)} & \textbf{(-0.053, -0.020)} & \textbf{(0.016, 0.043)} & \textbf{(-0.017, -0.003)} \\ 
  Caries 9yr Occurrence o & \textbf{(-0.116, -0.067)} & (-0.001, 0.018) & (-0.003, 0.000) & (-0.004, 0.004) & \textbf{(0.005, 0.031)} & (-0.024, 0.002) & (-0.008, 0.007) \\ 
  Caries 9yr Occurrence od & \textbf{(0.014, 0.063)} & \textbf{(-0.025, -0.006)} & (-0.003, 0.000) & (-0.007, 0.002) & \textbf{(0.005, 0.029)} & \textbf{(-0.026, -0.004)} & (-0.001, 0.013) \\ 
  Caries 9yr Occurrence om & \textbf{(0.020, 0.079)} & (-0.003, 0.018) & (-0.002, 0.001) & (-0.008, 0.002) & (-0.031, 0.000) & \textbf{(0.002, 0.029)} & (-0.011, 0.004) \\ 
  Caries 13yr Occurrence b & \textbf{(-0.171, -0.016)} & (-0.012, 0.017) & (-0.000, 0.004) & (-0.004, 0.009) & (-0.037, 0.012) & (-0.011, 0.032) & (-0.005, 0.017) \\ 
  Caries 13yr Occurrence d & \textbf{(-0.333, -0.172)} & (-0.029, 0.005) & (-0.002, 0.003) & \textbf{(0.002, 0.014)} & \textbf{(-0.061, -0.002)} & \textbf{(0.001, 0.047)} & \textbf{(0.008, 0.032)} \\ 
  Caries 13yr Occurrence l & (-0.070, 0.118) & (-0.030, 0.002) & (-0.005, 0.001) & (-0.008, 0.006) & \textbf{(-0.094, -0.040)} & \textbf{(0.026, 0.073)} & (-0.006, 0.018) \\ 
  Caries 13yr Occurrence m & \textbf{(-0.176, -0.021)} & \textbf{(0.004, 0.038)} & (-0.002, 0.002) & (-0.003, 0.009) & (-0.041, 0.017) & (-0.004, 0.043) & (-0.012, 0.011) \\ 
  Caries 13yr Occurrence o & (-0.123, 0.046) & (-0.024, 0.007) & \textbf{(0.000, 0.005)} & (-0.010, 0.004) & (-0.003, 0.056) & (-0.034, 0.013) & (-0.016, 0.009) \\ 
  Caries 13yr Occurrence od & (-0.069, 0.083) & \textbf{(-0.048, -0.019)} & (-0.004, 0.000) & (-0.005, 0.009) & \textbf{(-0.071, -0.015)} & \textbf{(0.012, 0.056)} & \textbf{(0.005, 0.026)} \\ 
  Caries 13yr Occurrence om & (-0.163, -0.005) & (-0.014, 0.017) & (-0.003, 0.001) & \textbf{(0.004, 0.017)} & (-0.041, 0.013) & (-0.000, 0.039) & \textbf{(0.002, 0.024)} \\ 
  Caries 17yr Occurrence b & (-0.070, 0.049) & (-0.015, 0.004) & (-0.004, 0.001) & (-0.001, 0.014) & (-0.044, 0.002) & (-0.008, 0.012) & (-0.014, 0.015) \\
Caries 17yr Occurrence d & \textbf{(0.004, 0.129)} & (-0.010, 0.009) & \textbf{(0.002, 0.008)} & \textbf{(0.010, 0.027)} & \textbf{(-0.075, -0.027)} & (-0.010, 0.008) & \textbf{(-0.056, -0.026)} \\
Caries 17yr Occurrence l & \textbf{(0.069, 0.208)} & (-0.011, 0.008) & (-0.003, 0.002) & (-0.005, 0.013) & (-0.026, 0.023) & (-0.022, 0.001) & \textbf{(0.000, 0.035)} \\
Caries 17yr Occurrence m & (-0.013, 0.115) & (-0.015, 0.003) & (-0.003, 0.001) & (-0.014, 0.003) & (-0.007, 0.042) & \textbf{(-0.022, -0.000)} & (-0.028, 0.000) \\
Caries 17yr Occurrence o & \textbf{(-0.150, -0.027)} & (-0.014, 0.002) & (-0.002, 0.002) & (-0.005, 0.011) & (-0.038, 0.004) & (-0.005, 0.014) & (-0.025, 0.003) \\
Caries 17yr Occurrence od & \textbf{(0.036, 0.157)} & (-0.003, 0.014) & (-0.001, 0.004) & \textbf{(0.005, 0.022)} & (-0.045, 0.003) & (-0.021, 0.002) & (-0.015, 0.016) \\
Caries 17yr Occurrence om & (-0.000, 0.126) & (-0.019, 0.001) & (-0.001, 0.004) & \textbf{(0.005, 0.020)} & (-0.038, 0.006) & \textbf{(-0.022, -0.003)} & (-0.026, 0.004) \\
\addlinespace
\multicolumn{8}{l}{\textit{Severity component: disease severity among caries-affected teeth}} \\[2pt]
Caries 9yr Severity b & \textbf{(0.019, 0.095)} & \textbf{(0.002, 0.014)} & \textbf{(-0.005, -0.001)} & (-0.001, 0.009) & (-0.013, 0.007) & (-0.008, 0.006) & \textbf{(0.005, 0.016)} \\ 
  Caries 9yr Severity d & (-0.019, 0.054) & (-0.004, 0.008) & (-0.003, 0.001) & (-0.006, 0.003) & \textbf{(-0.030, -0.006)} & \textbf{(0.002, 0.014)} & (-0.001, 0.007) \\ 
  Caries 9yr Severity l & \textbf{(0.025, 0.119)} & (-0.002, 0.010) & (-0.003, 0.001) & (-0.002, 0.008) & (-0.024, 0.001) & (-0.005, 0.010) & (-0.001, 0.010) \\ 
  Caries 9yr Severity m & \textbf{(-0.086, -0.018)} & \textbf{(0.004, 0.016)} & \textbf{(-0.006, -0.002)} & (-0.001, 0.007) & \textbf{(-0.041, -0.016)} & (-0.001, 0.012) & \textbf{(-0.017, -0.003)} \\ 
  Caries 9yr Severity o & (-0.031, 0.046) & \textbf{(0.001, 0.013)} & (-0.002, 0.002) & (-0.001, 0.006) & (-0.006, 0.016) & (-0.007, 0.004) & (-0.003, 0.004) \\ 
  Caries 9yr Severity od & \textbf{(0.014, 0.047)} & \textbf{(0.002, 0.013)} & \textbf{(-0.006, -0.002)} & \textbf{(0.000, 0.008)} & \textbf{(-0.023, -0.002)} & (-0.006, 0.008) & \textbf{(0.003, 0.013)} \\ 
  Caries 9yr Severity om & (-0.040, 0.041) & (-0.012, 0.001) & \textbf{(0.001, 0.005)} & (-0.005, 0.005) & (-0.008, 0.012) & \textbf{(0.001, 0.016)} & (-0.007, 0.004) \\ 
  Caries 13yr Severity b & \textbf{(0.049, 0.154)} & \textbf{(0.016, 0.046)} & \textbf{(-0.003, -0.000)} & \textbf{(-0.025, -0.012)} & \textbf{(0.006, 0.025)} & \textbf{(-0.027, -0.012)} & \textbf{(-0.031, -0.013)} \\ 
  Caries 13yr Severity d & \textbf{(0.020, 0.126)} & \textbf{(0.016, 0.048)} & (-0.001, 0.001) & \textbf{(-0.020, -0.006)} & (-0.003, 0.017) & \textbf{(-0.017, -0.003)} & \textbf{(-0.021, -0.003)} \\ 
  Caries 13yr Severity l & (-0.035, 0.073) & \textbf{(0.022, 0.056)} & (-0.002, 0.000) & \textbf{(-0.020, -0.004)} & (-0.008, 0.013) & (-0.005, 0.010) & (-0.020, 0.001) \\ 
  Caries 13yr Severity m & \textbf{(-0.102, -0.005)} & (-0.000, 0.027) & \textbf{(0.000, 0.003)} & (-0.007, 0.008) & (-0.002, 0.019) & (-0.008, 0.007) & (-0.012, 0.004) \\ 
  Caries 13yr Severity o & (-0.037, 0.051) & (-0.003, 0.024) & (-0.001, 0.001) & \textbf{(-0.012, -0.000)} & (-0.010, 0.007) & (-0.006, 0.006) & (-0.007, 0.008) \\ 
  Caries 13yr Severity od & (-0.007, 0.081) & (-0.001, 0.022) & (-0.001, 0.001) & \textbf{(-0.012, -0.001)} & (-0.001, 0.020) & \textbf{(-0.016, -0.002)} & \textbf{(-0.020, -0.004)} \\ 
  Caries 13yr Severity om & (-0.008, 0.092) & (-0.018, 0.012) & (-0.001, 0.001) & (-0.016, 0.000) & (-0.019, 0.001) & \textbf{(-0.016, -0.002)} & (-0.013, 0.004) \\ 
  Caries 17yr Severity b & \textbf{(0.075, 0.310)} & \textbf{(0.014, 0.048)} & (-0.004, 0.002) & (-0.019, 0.011) & \textbf{(0.030, 0.093)} & (-0.022, 0.026) & (-0.048, 0.006) \\
Caries 17yr Severity d & (-0.069, 0.145) & (-0.008, 0.034) & \textbf{(0.001, 0.007)} & (-0.021, 0.010) & \textbf{(0.005, 0.065)} & (-0.025, 0.020) & \textbf{(-0.072, -0.002)} \\
Caries 17yr Severity l & (-0.160, 0.034) & (-0.013, 0.033) & (-0.000, 0.006) & (-0.027, 0.011) & (-0.016, 0.058) & (-0.038, 0.017) & \textbf{(-0.079, -0.000)} \\
Caries 17yr Severity m & (-0.106, 0.135) & (-0.007, 0.039) & (-0.003, 0.003) & (-0.032, 0.004) & (-0.022, 0.032) & (-0.013, 0.033) & (-0.046, 0.025) \\
Caries 17yr Severity o & (-0.047, 0.120) & (-0.027, 0.015) & (-0.002, 0.003) & (-0.013, 0.023) & (-0.024, 0.031) & (-0.038, 0.005) & (-0.031, 0.042) \\
Caries 17yr Severity od & (-0.073, 0.137) & (-0.006, 0.032) & (-0.002, 0.004) & (-0.014, 0.016) & (-0.027, 0.041) & (-0.001, 0.053) & (-0.033, 0.026) \\
Caries 17yr Severity om & (-0.073, 0.137) & (-0.013, 0.030) & (-0.004, 0.003) & (-0.032, 0.005) & (-0.045, 0.020) & (-0.040, 0.008) & (-0.063, 0.004) \\
  \bottomrule
\end{tabular}}
\label{tab:C17long}
\end{table}

\begin{table}[ht]
\centering
\caption{Longitudinal joint model: Wasserstein barycenter-based summarization by tooth type and examination age. Tooth types: incisor, canine, premolar, molar. Occurrence block (upper rows): odds of developing disease. Severity block (lower rows): disease severity among affected teeth. Results are shown for both fluorosis (tooth zones aggregated) and caries (tooth surfaces aggregated) at ages 9, 13, and 17 years. Predictors are time-varying, updated for each examination period. AUC = area under the curve for the relevant period. Intervals are 95\% posterior credible intervals; those excluding zero are in bold. \textit{Note: the number of assessable teeth increases across ages as the permanent dentition erupts. Only tooth-type--age combinations with observed data contribute to inference; the joint Tucker model borrows information across ages so that estimates for partially observed tooth types are stabilised by the shared time and subject factors.}}
\scriptsize
\setlength{\tabcolsep}{3pt}
\renewcommand{\arraystretch}{0.7}
\resizebox{\textwidth}{!}{
\begin{tabular}{lrrrrrrr}
\toprule
 & \shortstack{Dental\\Exam}   & \shortstack{AUC Fluoride\\Intake (mg)}   & \shortstack{AUC Soft Drink\\Intake (oz)}   & \shortstack{Tooth Brushing\\Freq/Day Avg}   & \shortstack{Dental Visits\\Past 6 mo Avg}   & \shortstack{Fluoride\\Treatment 6 mo Avg}   & \shortstack{Home Water\\Fluoride (ppm) Avg} \\ \\
\midrule
\multicolumn{8}{l}{\textit{Occurrence component: odds of developing fluorosis, by tooth type}} \\[2pt]
  Fluorosis 9yr Occurrence Incisor & (-0.032, 0.008) & (-0.011, 0.016) & \textbf{(0.000, 0.005)} & (-0.004, 0.006) & (-0.010, 0.014) & (-0.025, 0.004) & \textbf{(0.002, 0.012)} \\ 
  Fluorosis 9yr Occurrence Canine & (-0.040, 0.013) & (-0.017, 0.019) & \textbf{(0.000, 0.004)} & (-0.003, 0.006) & (-0.014, 0.016) & (-0.025, 0.015) & \textbf{(0.000, 0.017)} \\ 
  Fluorosis 9yr Occurrence Premolar & (-0.007, 0.037) & (-0.013, 0.012) & \textbf{(0.001, 0.005)} & (-0.006, 0.001) & (-0.009, 0.009) & (-0.018, 0.014) & \textbf{(0.012, 0.027)} \\ 
  Fluorosis 9yr Occurrence Molar & (-0.024, 0.016) & (-0.000, 0.022) & \textbf{(0.000, 0.003)} & (-0.006, 0.001) & (-0.020, 0.002) & (-0.009, 0.010) & (-0.005, 0.007) \\ 
  Fluorosis 13yr Occurrence Incisor & \textbf{(0.021, 0.070)} & \textbf{(0.007, 0.030)} & \textbf{(-0.006, -0.002)} & (-0.004, 0.005) & \textbf{(0.009, 0.047)} & \textbf{(-0.052, -0.015)} & (-0.022, 0.001) \\ 
  Fluorosis 13yr Occurrence Canine & (-0.029, 0.088) & (-0.003, 0.020) & \textbf{(0.001, 0.005)} & (-0.006, 0.015) & (-0.032, 0.015) & (-0.013, 0.032) & (-0.017, 0.006) \\ 
  Fluorosis 13yr Occurrence Premolar & \textbf{(0.022, 0.089)} & \textbf{(0.012, 0.033)} & (-0.000, 0.005) & (-0.000, 0.012) & \textbf{(0.028, 0.069)} & \textbf{(-0.066, -0.014)} & \textbf{(-0.023, -0.002)} \\ 
  Fluorosis 13yr Occurrence Molar & \textbf{(0.029, 0.117)} & (-0.001, 0.016) & (-0.001, 0.003) & (-0.003, 0.008) & \textbf{(0.008, 0.034)} & \textbf{(-0.035, -0.006)} & \textbf{(-0.017, -0.003)} \\ 
  Fluorosis 17yr Occurrence Incisor & (-0.030, 0.005) & \textbf{(0.001, 0.018)} & (-0.004, 0.000) & (-0.001, 0.010) & (-0.013, 0.008) & (-0.010, 0.006) & \textbf{(-0.024, -0.007)} \\ 
  Fluorosis 17yr Occurrence Canine & (-0.173, -0.009) & (-0.004, 0.012) & \textbf{(0.001, 0.007)} & (-0.004, 0.008) & (-0.019, 0.013) & (-0.011, 0.015) & \textbf{(0.000, 0.019)} \\ 
  Fluorosis 17yr Occurrence Premolar & (-0.109, -0.020) & \textbf{(0.013, 0.027)} & \textbf{(-0.004, -0.000)} & \textbf{(0.005, 0.019)} & (-0.012, 0.009) & (-0.002, 0.015) & (-0.004, 0.015) \\ 
  Fluorosis 17yr Occurrence Molar & (-0.039, 0.043) & (-0.002, 0.006) & (-0.002, 0.001) & (-0.009, 0.001) & (-0.010, 0.009) & (-0.003, 0.009) & (-0.001, 0.011) \\ 
\addlinespace
\multicolumn{8}{l}{\textit{Occurrence component: odds of developing caries, by tooth type}} \\[2pt]
  Caries 9yr Occurrence Incisor & \textbf{(-0.063, -0.011)} & (-0.001, 0.015) & (-0.001, 0.002) & (-0.001, 0.004) & (-0.006, 0.021) & (-0.015, 0.004) & (-0.006, 0.005) \\ 
  Caries 9yr Occurrence Canine & \textbf{(0.019, 0.073)} & \textbf{(-0.029, -0.011)} & (-0.001, 0.002) & (-0.007, 0.001) & \textbf{(0.005, 0.036)} & \textbf{(-0.026, -0.003)} & \textbf{(0.004, 0.017)} \\ 
  Caries 9yr Occurrence Premolar & (-0.020, 0.023) & (-0.007, 0.006) & (-0.001, 0.002) & (-0.004, 0.002) & (-0.006, 0.011) & (-0.013, 0.007) & (-0.001, 0.008) \\ 
  Caries 9yr Occurrence Molar & \textbf{(0.003, 0.042)} & \textbf{(0.007, 0.020)} & \textbf{(-0.005, -0.003)} & \textbf{(-0.010, -0.003)} & (-0.019, 0.002) & \textbf{(0.004, 0.021)} & (-0.009, 0.003) \\ 
  Caries 13yr Occurrence Incisor & (-0.125, 0.002) & (-0.023, 0.001) & (-0.001, 0.001) & (-0.004, 0.008) & (-0.039, 0.010) & (-0.005, 0.037) & (-0.005, 0.012) \\ 
  Caries 13yr Occurrence Canine & (-0.103, 0.024) & (-0.023, 0.003) & (-0.003, 0.001) & \textbf{(0.002, 0.017)} & \textbf{(-0.052, -0.007)} & \textbf{(0.001, 0.032)} & (-0.004, 0.015) \\ 
  Caries 13yr Occurrence Premolar & \textbf{(-0.119, -0.002)} & (-0.005, 0.017) & (-0.002, 0.002) & (-0.004, 0.005) & \textbf{(-0.049, -0.014)} & \textbf{(0.005, 0.031)} & (-0.007, 0.009) \\ 
  Caries 13yr Occurrence Molar & \textbf{(-0.174, -0.043)} & \textbf{(-0.019, -0.000)} & (-0.001, 0.002) & (-0.002, 0.008) & (-0.034, 0.003) & \textbf{(0.013, 0.040)} & \textbf{(0.011, 0.024)} \\ 
  Caries 17yr Occurrence Incisor & (-0.038, 0.079) & (-0.011, 0.004) & (0.000, 0.004) & \textbf{(0.001, 0.016)} & (-0.037, 0.002) & (-0.013, 0.006) & (-0.020, 0.001) \\
Caries 17yr Occurrence Canine & \textbf{(0.021, 0.127)} & (-0.009, 0.008) & (-0.001, 0.004) & \textbf{(0.005, 0.020)} & (-0.041, 0.002) & \textbf{(-0.023, -0.004)} & (-0.015, 0.006) \\
Caries 17yr Occurrence Premolar & (-0.000, 0.094) & (-0.007, 0.008) & (-0.002, 0.001) & (-0.009, 0.005) & (-0.018, 0.019) & (-0.007, 0.005) & (-0.010, 0.011) \\
Caries 17yr Occurrence Molar & \textbf{(0.006, 0.092)} & (-0.015, 0.000) & (-0.001, 0.002) & \textbf{(0.006, 0.017)} & \textbf{(-0.041, -0.007)} & \textbf{(-0.013, -0.001)} & \textbf{(-0.025, -0.007)} \\
\addlinespace
\multicolumn{8}{l}{\textit{Severity component: fluorosis severity by tooth type}} \\[2pt]
Fluorosis 9yr Severity Incisor & (-0.028, 0.005) & (-0.014, 0.002) & (-0.002, 0.002) & (-0.004, 0.002) & \textbf{(0.002, 0.008)} & \textbf{(0.002, 0.009)} & \textbf{(0.002, 0.011)} \\ 
  Fluorosis 9yr Severity Canine & (-0.003, 0.059) & (-0.006, 0.006) & \textbf{(0.000, 0.003)} & (-0.001, 0.005) & \textbf{(0.006, 0.016)} & (-0.001, 0.007) & \textbf{(0.002, 0.008)} \\ 
  Fluorosis 9yr Severity Premolar & \textbf{(0.008, 0.036)} & (-0.006, 0.003) & \textbf{(0.001, 0.003)} & \textbf{(0.002, 0.006)} & (-0.000, 0.007) & \textbf{(0.002, 0.010)} & (-0.004, 0.004) \\ 
  Fluorosis 9yr Severity Molar & \textbf{(0.010, 0.042)} & (-0.005, 0.002) & (-0.000, 0.002) & (-0.001, 0.003) & \textbf{(0.003, 0.011)} & \textbf{(-0.005, -0.001)} & (-0.004, 0.002) \\ 
  Fluorosis 13yr Severity Incisor & \textbf{(0.021, 0.070)} & (-0.009, 0.004) & (-0.001, 0.001) & \textbf{(-0.007, -0.000)} & \textbf{(-0.017, -0.005)} & \textbf{(-0.022, -0.007)} & \textbf{(-0.027, -0.009)} \\ 
  Fluorosis 13yr Severity Canine & \textbf{(-0.115, -0.021)} & (-0.007, 0.012) & (-0.001, 0.001) & \textbf{(0.006, 0.015)} & \textbf{(0.000, 0.015)} & (-0.013, 0.010) & (-0.017, 0.006) \\ 
  Fluorosis 13yr Severity Premolar & \textbf{(0.022, 0.089)} & (-0.004, 0.015) & (-0.001, 0.001) & (-0.005, 0.003) & (-0.001, 0.016) & (-0.010, 0.008) & (-0.014, 0.008) \\ 
  Fluorosis 13yr Severity Molar & (-0.014, 0.044) & \textbf{(-0.014, -0.003)} & (-0.001, 0.001) & (-0.002, 0.005) & \textbf{(0.000, 0.009)} & \textbf{(-0.015, -0.001)} & \textbf{(0.010, 0.026)} \\ 
  Fluorosis 17yr Severity Incisor & (-0.030, 0.020) & (-0.005, 0.006) & (-0.001, 0.000) & (-0.006, 0.003) & (-0.008, 0.005) & (-0.012, 0.004) & (-0.005, 0.012) \\ 
  Fluorosis 17yr Severity Canine & (-0.031, 0.027) & \textbf{(0.000, 0.014)} & \textbf{(-0.003, -0.002)} & \textbf{(-0.022, -0.011)} & (-0.000, 0.018) & \textbf{(-0.032, -0.013)} & \textbf{(-0.023, -0.004)} \\ 
  Fluorosis 17yr Severity Premolar & (-0.021, 0.044) & (-0.010, 0.002) & \textbf{(-0.002, -0.000)} & (-0.009, 0.002) & \textbf{(0.001, 0.022)} & (-0.014, 0.000) & (-0.006, 0.010) \\ 
  Fluorosis 17yr Severity Molar & \textbf{(-0.039, -0.000)} & (-0.006, 0.004) & (-0.001, 0.001) & (-0.001, 0.008) & (-0.007, 0.007) & \textbf{(0.012, 0.030)} & (-0.016, 0.000) \\ 
\addlinespace
\multicolumn{8}{l}{\textit{Severity component: caries severity by tooth type}} \\[2pt]
Caries 9yr Severity Incisor & (-0.000, 0.046) & (-0.008, 0.001) & (-0.002, 0.001) & (-0.003, 0.003) & (-0.013, 0.002) & (-0.001, 0.008) & \textbf{(0.003, 0.009)} \\ 
  Caries 9yr Severity Canine & (-0.011, 0.064) & \textbf{(0.003, 0.017)} & \textbf{(-0.007, -0.002)} & \textbf{(0.006, 0.016)} & \textbf{(-0.026, -0.004)} & (-0.009, 0.013) & (-0.006, 0.008) \\ 
  Caries 9yr Severity Premolar & (-0.023, 0.047) & \textbf{(0.007, 0.016)} & (-0.002, 0.001) & (-0.004, 0.004) & (-0.015, 0.001) & (-0.005, 0.004) & (-0.004, 0.003) \\ 
  Caries 9yr Severity Molar & (-0.007, 0.028) & (-0.000, 0.007) & \textbf{(-0.003, -0.000)} & \textbf{(0.000, 0.004)} & \textbf{(-0.020, -0.005)} & \textbf{(0.001, 0.012)} & \textbf{(0.002, 0.008)} \\ 
  Caries 13yr Severity Incisor & \textbf{(0.037, 0.137)} & \textbf{(0.009, 0.030)} & (-0.002, 0.000) & \textbf{(-0.018, -0.008)} & (-0.001, 0.015) & \textbf{(-0.017, -0.006)} & \textbf{(-0.020, -0.007)} \\ 
  Caries 13yr Severity Canine & (-0.099, 0.023) & \textbf{(0.008, 0.028)} & (-0.001, 0.001) & \textbf{(-0.019, -0.004)} & (-0.001, 0.014) & (-0.015, 0.003) & (-0.016, 0.002) \\ 
  Caries 13yr Severity Premolar & (-0.056, 0.025) & \textbf{(0.016, 0.037)} & (-0.000, 0.001) & (-0.008, 0.001) & (-0.006, 0.008) & \textbf{(0.001, 0.011)} & (-0.007, 0.006) \\ 
  Caries 13yr Severity Molar & \textbf{(0.013, 0.098)} & \textbf{(0.005, 0.021)} & (-0.001, 0.001) & \textbf{(-0.013, -0.006)} & (-0.001, 0.011) & \textbf{(-0.017, -0.007)} & \textbf{(-0.018, -0.008)} \\ 
  Caries 17yr Severity Incisor & (-0.004, 0.190) & \textbf{(0.000, 0.029)} & (-0.001, 0.004) & (-0.013, 0.011) & (-0.009, 0.036) & (-0.005, 0.026) & \textbf{(-0.050, -0.003)} \\
Caries 17yr Severity Canine & (-0.058, 0.125) & (-0.026, 0.012) & (-0.003, 0.003) & (-0.016, 0.018) & (-0.037, 0.029) & (-0.033, 0.018) & (-0.036, 0.037) \\
Caries 17yr Severity Premolar & (-0.093, 0.048) & (-0.007, 0.021) & (-0.001, 0.003) & (-0.019, 0.009) & \textbf{(0.007, 0.056)} & \textbf{(-0.038, -0.006)} & (-0.040, 0.016) \\
Caries 17yr Severity Molar & (-0.013, 0.110) & \textbf{(0.008, 0.037)} & (-0.001, 0.002) & \textbf{(-0.023, -0.002)} & \textbf{(0.000, 0.035)} & (-0.009, 0.021) & \textbf{(-0.047, -0.009)} \\
 \bottomrule
\end{tabular}}
\label{tab:type_age}
\end{table}

\paragraph*{Fluorosis (zone $\times$ age barycenters)}
Tables~\ref{tab:F17long} and \ref{tab:type_age} together show substantial heterogeneity in fluorosis associations
across examination ages and tooth zones and tooth types. Several broad patterns emerge in the
\emph{Occurrence} block. At age 9, soft drink intake and home water fluoride show
positive occurrence effects in selected zones, indicating increased odds of
developing fluorosis. At ages 13 and 17, fluoride intake measured over the
period preceding each examination
becomes more consistently positive in the occurrence component, particularly
for cervical and incisal zones and for premolars and molars, again indicating
increased risk of developing fluorosis \citep{Kang2023,sarkar2024analyzing}.

In the \emph{Severity} block, fluoride-related predictors show age- and
zone-specific patterns. At age 9, dental visits show positive severity effects
across zones, while at age 13 some severity effects of fluoride treatment
become negative (lower severity). These patterns indicate that fluoride-related
predictors remain important, but their estimated effects are not constant across
developmental stages or tooth zones, which is biologically plausible given
tooth-specific enamel formation windows \citep{Kang2023,sarkar2024analyzing}.

\paragraph*{Caries (surface $\times$ age barycenters)}
Tables~\ref{tab:C17long} and \ref{tab:type_age} show that caries associations evolve with age and differ by
tooth surface. In the \emph{Occurrence} block, fluoride-related variables
(particularly fluoride treatment) show positive occurrence effects at several
ages and surfaces, again likely reflecting reactive administration to
higher-risk children. Soft drink intake shows positive occurrence effects for
selected surfaces at age 13. Home water fluoride shows negative occurrence
effects on some surfaces at age 9, consistent with a protective role.

In the \emph{Severity} block, fluoride intake AUC shows positive severity
effects for several surfaces at ages 13 and 17, which is unexpected and may
reflect confounding by baseline disease burden. Fluoride treatment and home
water fluoride show negative severity effects at ages 13 and 17 for several
tooth types (incisors, canines, molars), supporting a protective role in
limiting disease progression. Dental visits show positive severity effects
at some ages, again consistent with reactive care-seeking.

The longitudinal analysis highlights three main features. First, separating
disease occurrence from severity remains essential, since the two components
often display distinct and sometimes opposing patterns. Second, fluoride-related
predictors and soft drink intake remain central for both caries and fluorosis, but
their estimated effects are strongly moderated by age, zone/surface, and
tooth type. Third, the results confirm substantial spatial and developmental
heterogeneity, with some effects appearing only for specific tooth locations or
age groups.

\begin{figure}[htbp]
    \centering
    \includegraphics[width=0.85\textwidth, trim=0cm 0cm 0cm 1.5cm, clip=true]{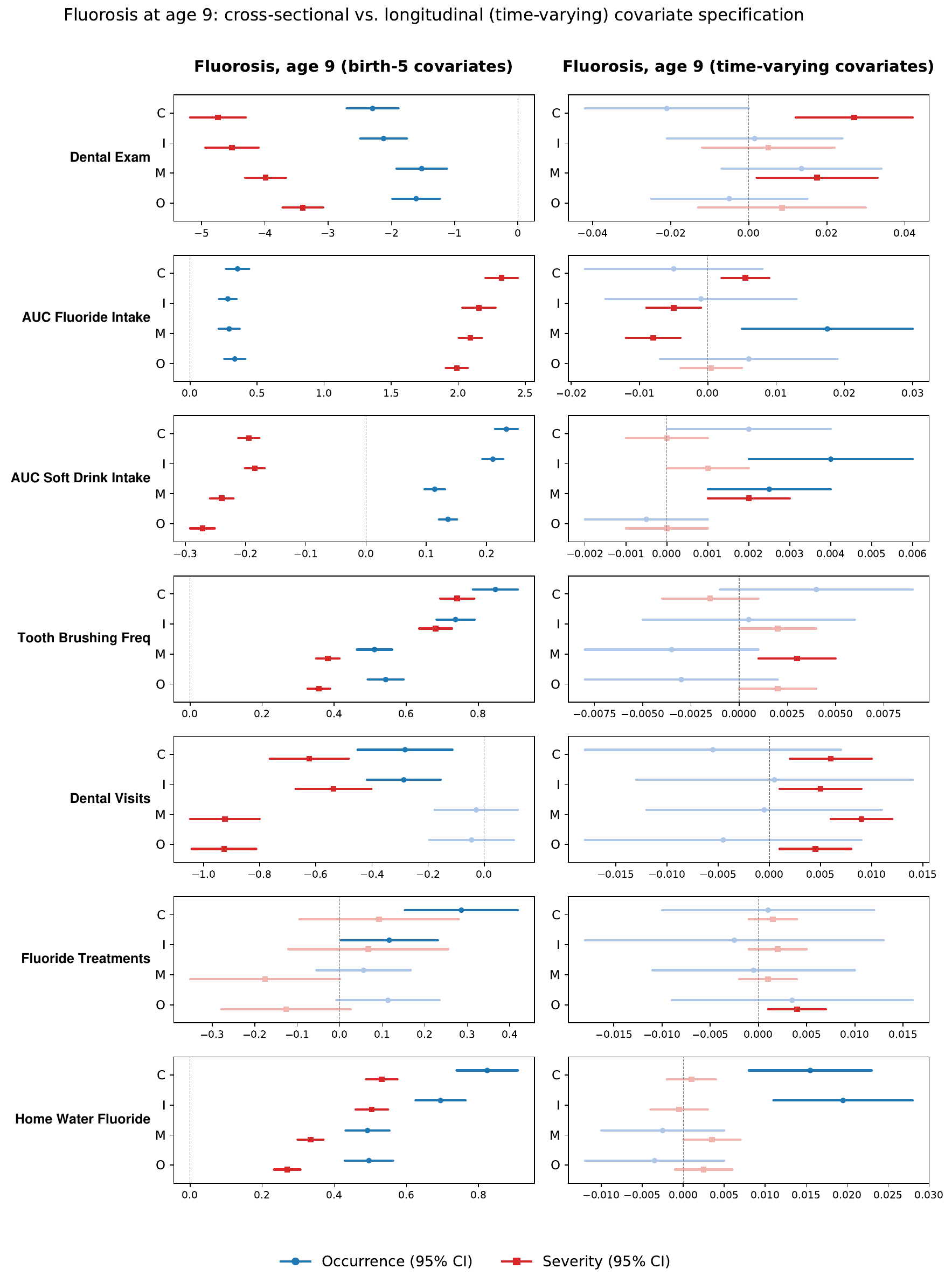}
    \caption{Fluorosis at age 9: model with birth-to-5 covariates (left panel, Table~\ref{tab:F9}) vs.\ model with age-5-to-9 covariates (right panel, age-9 rows of Table~\ref{tab:F17long}), for all 7 predictors. Occurrence (blue circle) and Severity (red square) markers show the posterior mean with 95\% credible interval; the darker/solid marker indicates the interval excludes zero.}
    \label{fig:fluorosis_age9_table3}
\end{figure}

\paragraph*{Cross-sectional versus time-varying covariate specification (Figure~\ref{fig:fluorosis_age9_table3})}
The two fluorosis analyses above use age-9 outcomes but differ in what the
predictors represent: Table~\ref{tab:F9} regresses fluorosis on covariates
accumulated from birth to age 5, whereas the age-9 rows of
Table~\ref{tab:F17long} regress the same age-9 outcome on covariates measured
only over the 
immediately preceding age-5-to-9 window.
Figure~\ref{fig:fluorosis_age9_table3} places these two specifications
side by side and makes the contrast explicit. Under the birth-to-5
specification (left panel), nearly every predictor shows a non-zero effect
across all four zones, for both the Occurrence and Severity components, with
credible intervals that are wide in absolute magnitude but consistently
bounded away from zero. Under the age-5-to-9 specification (right
panel), the great majority of intervals cross zero and the point estimates
themselves are roughly two orders of magnitude smaller; only a handful of
associations remain distinguishable from the null, most visibly for soft drink
intake and home water fluoride in the Occurrence component and for a subset of
zone--age combinations in the Severity component. This attenuation is
consistent with fluorosis being a developmental condition whose presentation
is set primarily during the enamel-formation window in early childhood
\citep{Levy2003,Kang2023,sarkar2024analyzing}: once the permanent incisors and
first molars have mineralized, exposures accrued afterward have comparatively
little further opportunity to alter fluorosis presentation. The comparison
therefore clarifies the complementary roles of the two specifications---the
cross-sectional, birth-to-5 model isolates the dominant early-life exposure
window that drives most of the identifiable fluorosis signal, while the
time-varying model shows that only a narrow set of behavioral and
environmental factors continue to register an effect once that window has
closed.

\begin{figure}[htbp]
    \centering
    \includegraphics[width=0.8\textwidth, trim=0cm 0cm 0cm 2cm, clip=true]{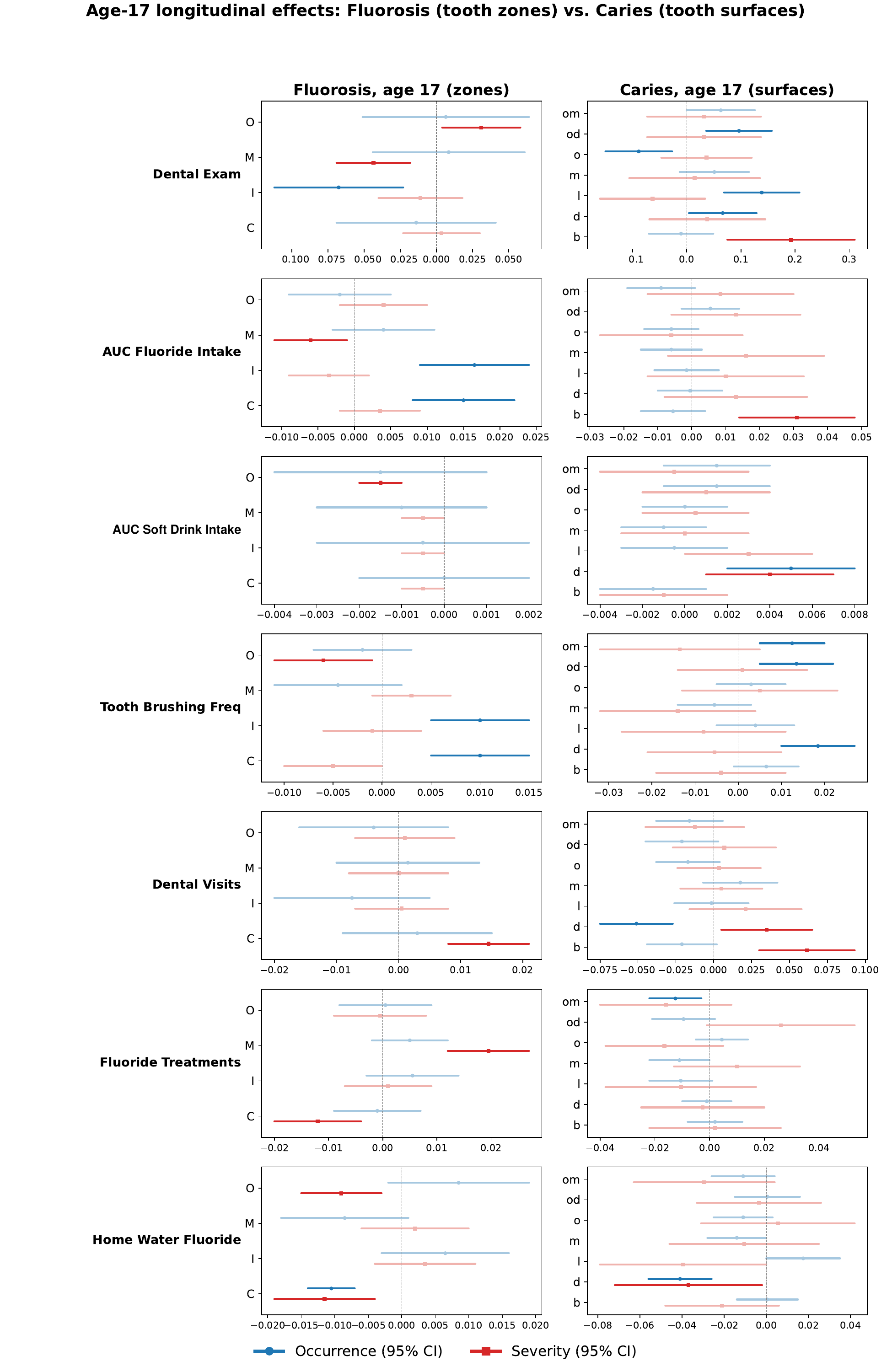}
    \caption{Age-17 longitudinal effects: Fluorosis by tooth zone (left panel) vs.\ Caries by tooth surface (right panel), for all 7 predictors. Occurrence (blue circle) and Severity (red square) markers show the posterior mean with 95\% credible interval; the darker/solid marker indicates the interval excludes zero. Summarizes the age-17 rows of Tables~\ref{tab:F17long} and \ref{tab:C17long}.}
    \label{fig:age17_zone_surface}
\end{figure}


\begin{figure}[htbp]
    \centering
    \includegraphics[width=0.85\textwidth, trim=0cm 0cm 0cm 2cm, clip=true]{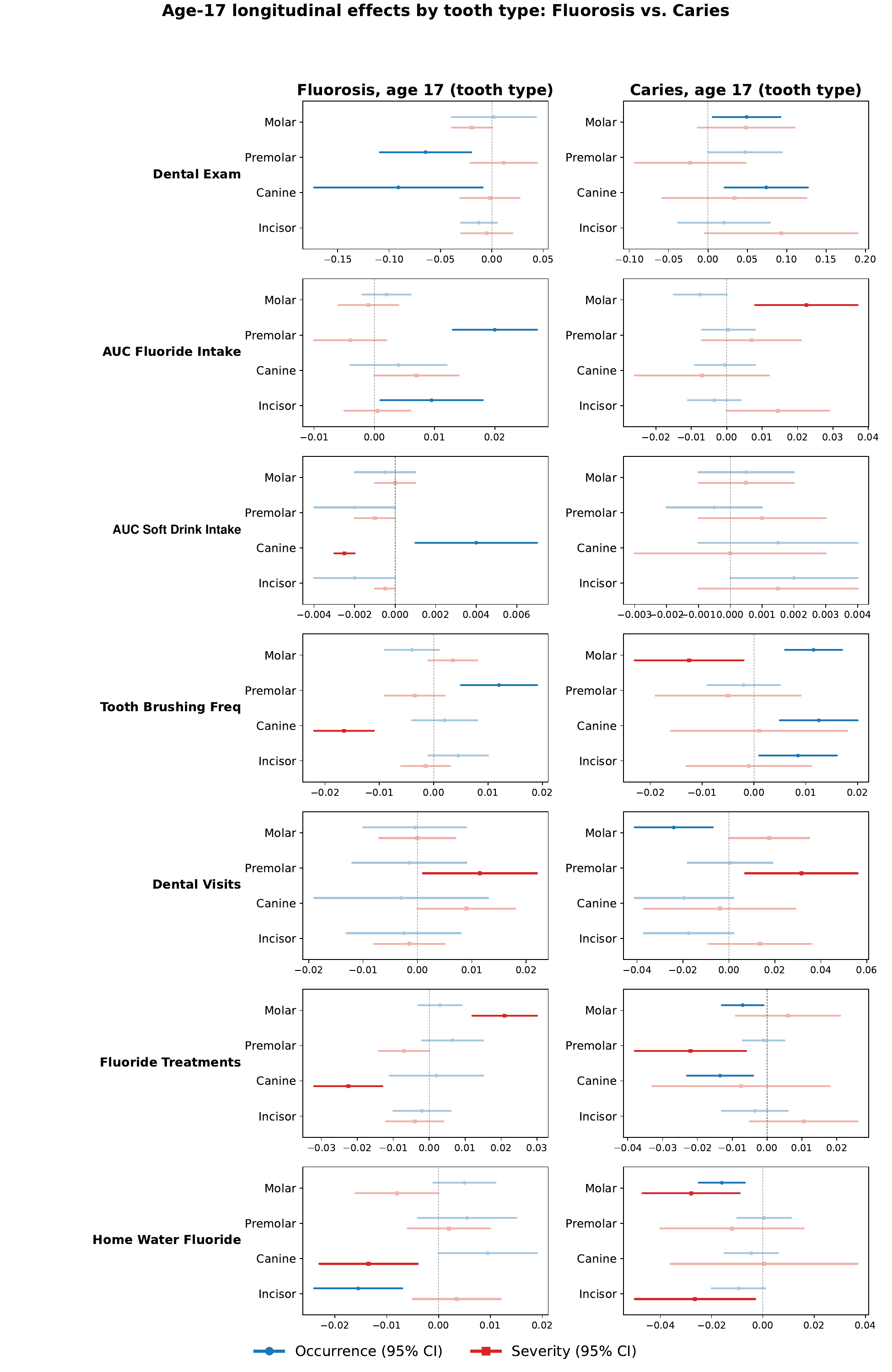}
    \caption{Age-17 longitudinal effects by tooth type: Fluorosis (left panel) vs.\ Caries (right panel), for all 7 predictors. Occurrence (blue circle) and Severity (red square) markers show the posterior mean with 95\% credible interval; the darker/solid marker indicates the interval excludes zero. Summarizes the age-17 rows of Table~\ref{tab:type_age}.}
    \label{fig:age17_toothtype}
\end{figure}


\subsection{\label{subsec:subpop} Sub-Population Inference at Age 17: Partitioning by Age-13 Caries Status}

The projection-based inference allows sub-population comparisons by
conditioning on earlier disease status \citep{Li2002,Broffitt2013}.
We focus this analysis on fluorosis at age 17 stratified by caries status at age 13
for two reasons. First, caries and fluorosis share early-life fluoride exposures as a
common determinant, so prior caries history is a plausible marker of the overall
fluoride and dietary exposure profile that also shapes fluorosis severity in later
adolescence. Second, the linked Tucker factorization couples the two outcomes
through shared subject factors, which means that a child's caries trajectory is
informative about their underlying susceptibility to fluorosis; stratifying by
age-13 caries status therefore makes substantive use of the joint model structure.
Table~\ref{tab:17yr} summarizes fluorosis
occurrence and severity at age 17 separately for children with and without
caries at age 13, providing insight into whether fluorosis risk profiles differ
between these dental sub-populations.

\textbf{Occurrence component.}
For children with \emph{no caries at age 13}, the occurrence component shows
positive effects of dental exam status across all four tooth zones,
indicating higher odds of developing fluorosis among those with more dental
contact (plausibly because fluoride-related exposures, including topical
applications, are higher in this group). Early cumulative fluoride intake shows
negative occurrence effects in the cervical, incisal, and middle zones,
indicating \emph{lower} odds of developing fluorosis in that sub-population---a
pattern that may reflect complex fluoride exposure histories or correlation
with other health-promoting behaviors. Fluoride treatment shows positive
occurrence effects in the cervical, incisal, and middle zones, and
tooth-brushing frequency shows negative effects in the cervical and incisal
zones.

For children with \emph{caries present at age 13}, the occurrence component
shows fewer non-zero associations. Dental exam status is negative in the
middle zone, indicating a lower probability of developing fluorosis in
this subgroup with that pattern. Fluoride intake is positive in the middle
zone, and fluoride treatment is negative in the incisal and middle zones.

\textbf{Severity component.}
Among children with \emph{no caries at age 13}, fluorosis severity at age 17
shows negative associations with dental exam status in the cervical and
occlusal zones, while fluoride treatment is negative in the middle zone. Overall, severity effects in the lower-risk group are modest and
localized.

Among children with \emph{caries present at age 13}, fluorosis severity
exhibits more heterogeneous patterns. Dental exam status is positive in the
middle zone. Fluoride intake is negative in the cervical and occlusal
zones, while it is weakly positive in the middle. Fluoride treatment
shows opposite patterns across zones, with a positive effect in the
cervical and a negative effect in the middle. Home water fluoride shows
positive effects in the cervical and occlusal zones.

\textbf{Comparison of sub-populations.}
Overall, the results show meaningful heterogeneity between the two
sub-populations. Children without caries at age 13 display clearer occurrence-
component associations---especially for dental exam status, early fluoride
intake, and fluoride treatment---while children with prior caries exhibit a
more mixed severity profile and fewer strong occurrence-component effects.
These findings suggest that fluorosis patterns at age 17 differ according to
earlier caries history \citep{Li2002}, and demonstrate that covariate effects are not
homogeneous across sub-populations defined by prior caries status \citep{Kang2021,sarkar2024analyzing}.

\begin{table}[ht]
\caption{Sub-population analysis: Wasserstein barycenter-based summarization for fluorosis at age 17, stratified by presence or absence of caries at age 13, by tooth zone. Occurrence block (upper rows): odds of developing fluorosis at age 17. Severity block (lower rows): fluorosis severity among affected teeth. Tooth zones: C = cervical, I = incisal, M = middle, O = occlusal. Intervals are 95\% posterior credible intervals; those excluding zero are in bold.}
\scriptsize
\setlength{\tabcolsep}{3pt}
\renewcommand{\arraystretch}{0.5}
\resizebox{\textwidth}{!}{
\begin{tabular}{lrrrrrrr}
\toprule
 & \shortstack{Dental\\Exam}   & \shortstack{AUC Fluoride\\Intake (mg)}   & \shortstack{AUC Soft Drink\\Intake (oz)}   & \shortstack{Tooth Brushing\\Freq/Day Avg}   & \shortstack{Dental Visits\\Past 6 mo Avg}   & \shortstack{Fluoride\\Treatment 6 mo Avg}   & \shortstack{Home Water\\Fluoride (ppm) Avg} \\ \\
\midrule
\multicolumn{8}{l}{\textit{Occurrence component: odds of developing fluorosis at age 17}} \\[2pt]
  Fluorosis 17yr Occurrence (No caries at age 13) C & \textbf{(0.001, 0.048)} & \textbf{(-0.018, -0.005)} & (-0.000, 0.001) & \textbf{(-0.014, -0.004)} & (-0.005, 0.011) & \textbf{(0.003, 0.012)} & (-0.003, 0.009) \\ 
  Fluorosis 17yr Occurrence (No caries at age 13) I & \textbf{(0.020, 0.079)} & \textbf{(-0.024, -0.009)} & \textbf{(0.000, 0.001)} & \textbf{(-0.013, -0.003)} & (-0.013, 0.007) & \textbf{(0.004, 0.014)} & (-0.003, 0.008) \\ 
  Fluorosis 17yr Occurrence (No caries at age 13) M & \textbf{(0.012, 0.072)} & \textbf{(-0.021, -0.008)} & (-0.000, 0.001) & (-0.000, 0.011) & (-0.016, 0.002) & \textbf{(0.000, 0.010)} & (-0.000, 0.012) \\ 
  Fluorosis 17yr Occurrence (No caries at age 13) O & \textbf{(0.005, 0.063)} & (-0.006, 0.009) & (-0.000, 0.001) & (-0.001, 0.008) & (-0.007, 0.009) & (-0.007, 0.003) & (-0.003, 0.009) \\ 
  Fluorosis 17yr Occurrence (Caries present at age 13) C & (-0.057, 0.040) & (-0.012, 0.005) & (-0.002, 0.001) & (-0.004, 0.002) & (-0.020, 0.007) & (-0.015, 0.001) & (-0.010, 0.013) \\ 
  Fluorosis 17yr Occurrence (Caries present at age 13) I & (-0.059, 0.054) & (-0.009, 0.009) & (-0.002, 0.001) & (-0.005, 0.002) & (-0.003, 0.024) & \textbf{(-0.023, -0.006)} & (-0.020, 0.002) \\ 
  Fluorosis 17yr Occurrence (Caries present at age 13) M & \textbf{(-0.096, -0.004)} & \textbf{(0.001, 0.020)} & (-0.001, 0.002) & (-0.004, 0.002) & (-0.009, 0.020) & \textbf{(-0.019, -0.003)} & (-0.010, 0.016) \\ 
  Fluorosis 17yr Occurrence (Caries present at age 13) O & (-0.087, 0.007) & (-0.008, 0.010) & (-0.000, 0.004) & (-0.005, 0.002) & (-0.010, 0.015) & (-0.007, 0.010) & (-0.025, 0.003) \\ 
\addlinespace
\multicolumn{8}{l}{\textit{Severity component: fluorosis severity at age 17}} \\[2pt]
  Fluorosis 17yr Severity (No caries at age 13) C & \textbf{(-0.047, -0.009)} & (-0.001, 0.006) & \textbf{(0.000, 0.001)} & (-0.000, 0.004) & (-0.009, 0.001) & (-0.005, 0.003) & (-0.007, 0.003) \\ 
  Fluorosis 17yr Severity (No caries at age 13) I & (-0.032, 0.007) & (-0.004, 0.004) & (-0.000, 0.000) & (-0.001, 0.004) & (-0.002, 0.008) & (-0.008, 0.000) & (-0.006, 0.003) \\ 
  Fluorosis 17yr Severity (No caries at age 13) M & (-0.032, 0.006) & (-0.002, 0.004) & \textbf{(-0.001, -0.000)} & (-0.001, 0.003) & (-0.011, 0.000) & \textbf{(-0.015, -0.007)} & (-0.004, 0.005) \\ 
  Fluorosis 17yr Severity (No caries at age 13) O & \textbf{(-0.056, -0.013)} & (-0.002, 0.004) & \textbf{(0.000, 0.001)} & \textbf{(0.001, 0.006)} & (-0.007, 0.006) & (-0.007, 0.003) & (-0.007, 0.004) \\ 
  Fluorosis 17yr Severity (Caries present at age 13) C & (-0.005, 0.050) & \textbf{(-0.011, -0.001)} & (-0.000, 0.001) & (-0.001, 0.007) & \textbf{(-0.016, -0.004)} & \textbf{(0.007, 0.019)} & \textbf{(0.006, 0.021)} \\ 
  Fluorosis 17yr Severity (Caries present at age 13) I & (-0.002, 0.049) & (-0.001, 0.007) & (-0.000, 0.001) & (-0.004, 0.003) & (-0.009, 0.003) & (-0.003, 0.009) & (-0.007, 0.004) \\ 
  Fluorosis 17yr Severity (Caries present at age 13) M & \textbf{(0.028, 0.087)} & \textbf{(0.000, 0.009)} & \textbf{(0.000, 0.001)} & \textbf{(-0.008, -0.001)} & (-0.002, 0.013) & \textbf{(-0.015, -0.002)} & (-0.008, 0.004) \\ 
  Fluorosis 17yr Severity (Caries present at age 13) O & (-0.026, 0.033) & \textbf{(-0.010, -0.000)} & \textbf{(0.001, 0.002)} & (-0.002, 0.006) & (-0.008, 0.007) & (-0.003, 0.009) & \textbf{(0.001, 0.014)} \\ 
  \bottomrule
\end{tabular}}
\label{tab:17yr}
\end{table}

\begin{figure}[htbp]
    \centering
    \includegraphics[width=0.8\textwidth, trim=0cm 0cm 0cm 2cm, clip=true]{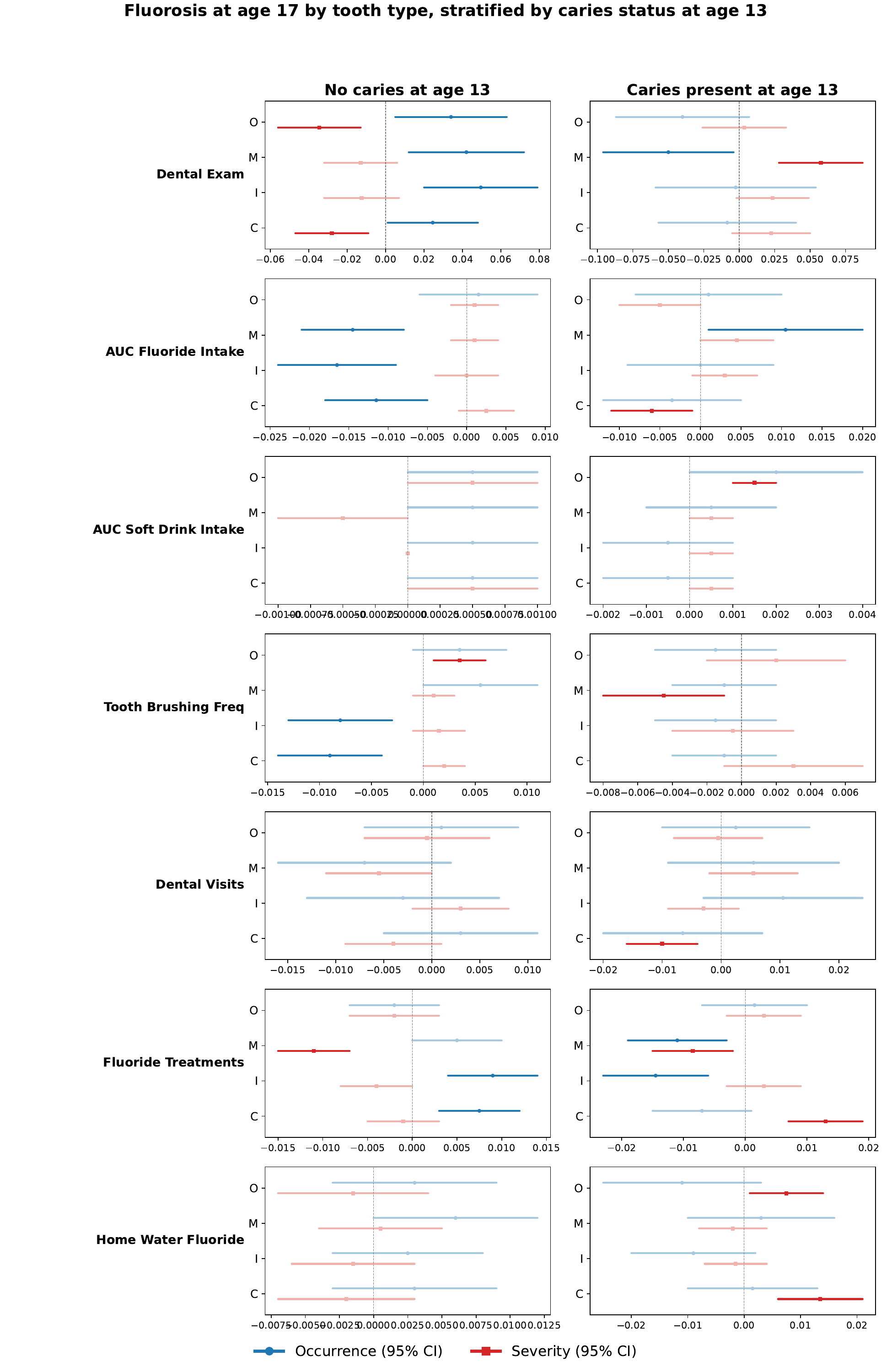}
    \caption{Fluorosis at age 17 by tooth zone, stratified by caries status at age 13: no caries at age 13 (left panel) vs.\ caries present at age 13 (right panel), for all 7 predictors. Occurrence (blue circle) and Severity (red square) markers show the posterior mean with 95\% credible interval; the darker/solid marker indicates the interval excludes zero. Summarizes Table~\ref{tab:17yr}.}
    \label{fig:table6_subpopulation}
\end{figure}


{\bf Comparison with Fluorosis- or Caries-Only Studies:} Several findings from our joint analysis agree with the earlier studies of each dental outcome. As found in \citet{sarkar2024analyzing}, fluoride intake and fluoride levels in home water are associated with the occurrence of fluorosis, although these associations differ by age and tooth location. Also, similar to the findings in \citet{mukherjee2024modeling}, soft drink intake is associated with a higher risk of caries at some ages and tooth surfaces, while tooth brushing, dental visits, professional fluoride treatment, and fluoride in home water are associated with lower caries risk or severity in several age--location groups. Our analysis also provides new findings. In particular, a predictor can have different associations with disease occurrence and severity, and these associations can vary across ages, tooth types, and surfaces or zones. By jointly modeling caries and fluorosis, we also find that the associations with fluorosis at age 17 differ depending on whether a child had caries at age 13. Children without caries at age 13 show clearer associations with the occurrence of fluorosis, whereas children with caries show more varied associations with fluorosis severity.

\section{\label{sec:simulation} Simulation Study}
 
We conduct simulation studies to evaluate the finite-sample performance of the
proposed linked Tucker hurdle--ordinal regression model and the associated
posterior inference procedure. To create a realistic but computationally
manageable setting, data are generated from a generalized linear mixed-effects
model for paired ordinal outcomes, with the structure chosen to mimic the main
features of the Iowa Fluoride Study. The fixed-effect component is specified
through a low-rank Tucker decomposition involving six active predictors,
corresponding to the covariates used in the real-data analysis except home
water fluoride. Subject-specific random effects are included to induce
within-subject dependence across repeated measurements and tooth locations.
 
The simulated datasets preserve several key features of the motivating
application, including paired caries and fluorosis outcomes, spatially-indexed
tooth locations, longitudinal measurements at three examination ages, and
substantial between-subject heterogeneity. To reduce computational burden while
retaining the essential spatial and longitudinal dependence structure, the
numbers of teeth, tooth surfaces, and tooth zones are reduced relative to the
full study. Specifically, four teeth are simulated for each subject, with four
spatial locations per tooth for both caries and fluorosis. The data-generating
model is intentionally not identical to the fitted model, particularly in the
way subject-specific effects are incorporated, allowing us to assess robustness
under moderate model misspecification.
 
For each simulated dataset, the proposed model is fit using all seven
covariates: the six active predictors used in data generation and one additional
noise predictor whose true coefficient tensor is identically zero. This setup
allows us to evaluate whether the procedure can recover truly associated
covariates while avoiding spurious findings for a null covariate. We consider
two sample sizes, $n=200$ and $n=300$, and generated 30 independent datasets for
each sample size.
 
Performance is summarized using location-specific variable-identification
accuracy. For each simulation replicate, spatial location \(q\), and
predictor, we construct the 95\% posterior credible interval for the
corresponding projected population-level coefficient, such as
\(\widetilde{\bmu}_q\), with the analogous projected coefficients used for
the other outcome and model components. Thus, each identification decision
is made separately for each \(q\) and predictor; it is not based on an
individualized \((i,q)\)-specific coefficient or on a Wasserstein-barycenter
summary across locations. A true positive is recorded when the interval for
an active predictor excludes zero, whereas a false positive is recorded when
the interval for the null predictor incorrectly excludes zero. These
quantities are computed separately for the ordinal-severity and occurrence
components and then summarized across the location-specific decisions and
simulation replicates.
 
Table~\ref{tab:sim_results} summarizes the simulation results. As expected,
the true-positive rate increases with sample size for both model components.
For the ordinal severity component, TP increases from 0.45 at $n=200$ to 0.71 at
$n=300$. For the occurrence component, TP increases from 0.57 to 0.82. The odds
component also shows lower FP rates than the ordinal component at both sample
sizes. Overall, the simulations suggest that
the linked Tucker structure improves recovery of the low-dimensional signal as
sample size increases, while also highlighting the difficulty of variable
selection in complex zero-inflated longitudinal ordinal data.

\begin{table}[ht]
\centering
\caption{Simulation results for sample sizes $n=200$ and $n=300$. TP denotes the empirical true-positive rate for the six active predictors, and FP denotes the empirical false-positive rate for the null predictor.}
\label{tab:sim_results}
\begin{tabular}{lccc}
\hline
Component & Sample size & TP & FP \\
\hline
Ordinal severity & $n=200$ & 0.45 & 0.29 \\
Occurrence (odds) & $n=200$ & 0.57 & 0.23 \\
\addlinespace
Ordinal severity & $n=300$ & 0.71 & 0.25 \\
Occurrence (odds) & $n=300$ & 0.82 & 0.19 \\
\hline
\end{tabular}
\end{table}

\paragraph*{Comparison with an unlinked model}
As one of the main methodological contributions of this paper is the shared
subject-mode factor linking the caries and fluorosis Tucker decompositions, we additionally compare the linked model
against an otherwise identical hurdle-ordinal Tucker model fit
\emph{separately} to each outcome, with no shared subject factor. Both models
are fit to the same simulated datasets, so that any difference in performance
is attributable to information sharing across outcomes rather than to
differences in data. The same ranks, priors, posterior settings, and TP/FP
definitions are used for both fits. The results are in Table~\ref{tab:sim_resultsunlinked}. Although the two models have similar TP and
FP rates for the occurrence component, the unlinked model performs worse on
the ordinal component, yielding lower TP and higher FP rates than the linked
model. This is expected because the severity model is estimated only from
nonzero responses and therefore has a smaller effective sample size; sharing
information across outcomes is consequently more beneficial for severity than
for occurrence coefficients.

\begin{table}[ht]
\centering
\caption{Simulation results for sample sizes $n=200$ and $n=300$ in the unlinked model. TP denotes the empirical true-positive rate for the six active predictors, and FP denotes the empirical false-positive rate for the null predictor.}
\label{tab:sim_resultsunlinked}
\begin{tabular}{lccc}
\hline
Component & Sample size & TP & FP \\
\hline
Ordinal severity & $n=200$ & 0.37 & 0.33 \\
Occurrence (odds) & $n=200$ & 0.55 & 0.25 \\
\addlinespace
Ordinal severity & $n=300$ & 0.66 & 0.30 \\
Occurrence (odds) & $n=300$ & 0.80 & 0.20 \\
\hline
\end{tabular}
\end{table}

\section{\label{sec:discussion} Discussion}

We proposed a joint hurdle--ordinal regression framework for paired,
zero-inflated ordinal outcomes with subject-specific, spatially varying, and
time-varying effects. A linked Tucker factorization provides the key structural
regularization, allowing dependence across outcomes through shared subject
factors while respecting the fact that the spatial resolutions for caries and
fluorosis are distinct. Throughout, we parameterize and present results so that
the \emph{occurrence component} models the odds of \emph{developing} disease
(positive effects indicate greater disease risk), and the \emph{severity component}
models disease burden among affected observations (positive effects indicate
greater severity). This unified sign convention---positive means more disease
in both components---ensures consistent interpretation across tables and
avoids the confusion that arises when one component models disease presence
and the other models disease absence.

Several extensions are natural. First, since outcomes are observed on a common
tooth-type grid, one could also share the tooth-type-level factors across the
two Tucker factorizations, which would impose additional structure on the joint
model. Second, one could explicitly decompose each coefficient tensor into a
global mean effect plus a low-rank heterogeneous component, which would sharpen the distinction between population-average and subject-specific effects.

The current projection-based summaries target interpretable marginal effects.
It is also possible to cluster subjects based on the shared subject-level
factors to identify sub-populations with similar disease trajectories; this
will be pursued in future work.

\section*{Data Availability}

The Iowa Fluoride Study data used in this analysis are available from the study
investigators upon reasonable request and subject to institutional (IRB)
approval, consistent with the human-subjects data-sharing restrictions under
which the cohort was recruited.

\section*{Acknowledgments}
The authors would like to thank Dr. Shoumi Sarkar (a former doctoral student under the last author) for helping us understand the data and for familiarizing us with the existing results. During the preparation of this work, the authors used generative AI to help polish the write-up. All content was subsequently reviewed and edited by the authors, who take full responsibility for the publication.

\section*{Funding}
This research was funded, in part, by NIH grant R03DE030502.

\bibliographystyle{plainnat}
\bibliography{references,bib,referencesN}

@article{hitchcock1927expression,
  title={The expression of a tensor or a polyadic as a sum of products},
  author={Hitchcock, Frank L},
  journal={Journal of Mathematics and Physics},
  volume={6},
  pages={164--189},
  year={1927},
  publisher={Wiley Online Library}
}

@article{tucker:1966,
  author={Tucker, L.},
  title={Some mathematical notes on three-mode factor analysis},
  journal={Psychometrika},
  volume={31},
  pages={273--282},
  year={1966}
}

@article{de_lathauwer_etal:2000,
  author={De Lathauwer, L. and De Moore, B. and Vandewalle, J.},
  title={A multilinear singular value decomposition},
  journal={SIAM Journal on Matrix Analysis and Applications},
  volume={21},
  pages={1253--1278},
  year={2000}
}

@article{kolda2009tensor,
  title={Tensor decompositions and applications},
  author={Kolda, Tamara G and Bader, Brett W},
  journal={SIAM Review},
  volume={51},
  number={3},
  pages={455--500},
  year={2009},
  publisher={SIAM}
}

@article{guhaniyogi2020bayesian,
  title={Bayesian methods for tensor regression},
  author={Guhaniyogi, Rajarshi},
  journal={Wiley StatsRef: Statistics Reference Online},
  pages={1--18},
  year={2020}
}

@article{agueh2011barycenters,
  author    = {Agueh, Martial and Carlier, Guillaume},
  title     = {Barycenters in the {W}asserstein Space},
  journal   = {SIAM Journal on Mathematical Analysis},
  year      = {2011},
  volume    = {43},
  number    = {2},
  pages     = {904--924},
  doi       = {10.1137/100805741}
}

@article{shi2023tensor,
  author={Shi, Y. and Shen, W.},
  title={Bayesian Methods in Tensor Analysis},
  journal={Statistics and Its Interface},
  volume={17},
  pages={249--274},
  year={2024}
}

@article{warren2001prevalence,
  title={Prevalence of dental fluorosis in the primary dentition},
  author={Warren, John J and Levy, Steven M and Kanellis, Michael J},
  journal={Journal of Public Health Dentistry},
  volume={61},
  number={2},
  pages={87--91},
  year={2001},
  publisher={Wiley Online Library}
}

@article{levy2001patterns,
  title={Patterns of fluoride intake from birth to 36 months},
  author={Levy, Steven M and Warren, John J and Davis, Charles S and Kirchner, H Lester and Kanellis, Michael J and Wefel, James S},
  journal={Journal of Public Health Dentistry},
  volume={61},
  number={2},
  pages={70--77},
  year={2001},
  publisher={Wiley Online Library}
}

@article{marshall2005roles,
  title={The roles of meal, snack, and daily total food and beverage exposures on caries experience in young children},
  author={Marshall, Teresa A and Broffitt, Barbara and Eichenberger-Gilmore, Julie and Warren, John J and Cunningham, Marsha A and Levy, Steven M},
  journal={Journal of Public Health Dentistry},
  volume={65},
  number={3},
  pages={166--173},
  year={2005},
  publisher={Wiley Online Library}
}

@article{warren2002dental,
  title={Dental caries in the primary dentition: assessing prevalence of cavitated and noncavitated lesions},
  author={Warren, John J and Levy, Steven M and Kanellis, Michael J},
  journal={Journal of Public Health Dentistry},
  volume={62},
  number={2},
  pages={109--114},
  year={2002},
  publisher={Wiley Online Library}
}

@article{marshall2005diet,
  title={Diet quality in young children is influenced by beverage consumption},
  author={Marshall, Teresa A and Eichenberger Gilmore, Julie M and Broffitt, Barbara and Stumbo, Phyllis J and Levy, Steven M},
  journal={Journal of the American College of Nutrition},
  volume={24},
  number={1},
  pages={65--75},
  year={2005},
  publisher={Taylor \& Francis}
}

@incollection{warren2021measurement,
  title={Measurement and Distribution of Dental Fluorosis},
  author={Warren, John J and Levy, Steven M},
  booktitle={Burt and Eklund's Dentistry, Dental Practice, and the Community},
  pages={218--226},
  year={2021},
  publisher={Elsevier}
}

@article{curtis2020decline,
  title={Decline in dental fluorosis severity during adolescence: a cohort study},
  author={Curtis, AM and Levy, SM and Cavanaugh, JE and Warren, JJ and Kolker, JL and Weber-Gasparoni, K},
  journal={Journal of dental research},
  volume={99},
  number={4},
  pages={388--394},
  year={2020},
  publisher={SAGE Publications Sage CA: Los Angeles, CA}
}

@article{yazdanbakhsh2024community,
  title={Community water fluoride cessation and rate of caries-related pediatric dental treatments under general anesthesia in Alberta, Canada},
  author={Yazdanbakhsh, Elnaz and Bohlouli, Babak and Patterson, Steven and Amin, Maryam},
  journal={Canadian Journal of Public Health},
  volume={115},
  number={2},
  pages={305--314},
  year={2024},
  publisher={Springer}
}

@article{Pitt2006,
author    = {Pitt, Michael and Chan, David and Kohn, Robert},
title     = {{Efficient Bayesian inference for Gaussian copula regression models}},
journal   = {Biometrika},
volume    = {93},
number    = {3},
pages     = {537--554},
year      = {2006}
}

@article{Kang2021,
author    = {Kang, Tong and Gaskins, Jeremy and , Steven and Datta, Somnath},
title     = {{A longitudinal Bayesian mixed effects model with hurdle Conway-Maxwell-Poisson distribution}},
journal   = {Statistics in Medicine},
volume    = {40},
number    = {6},
pages     = {1336--1356},
year      = {2021}
}

@article{Choo-Wosoba2016,
author    = {Choo-Wosoba, Hyoyoung and Levy, Steven M. and Datta, Somnath},
title     = {{Marginal regression models for clustered count data based on zero-inflated Conway-Maxwell-Poisson distribution with applications}},
journal   = {Biometrics},
volume    = {72},
number    = {2},
pages     = {606--618},
year      = {2016}
}

@article{Marshall2003,
author    = {Marshall, Teresa A. and Levy, Steven M. and Broffitt, Barbara and Warren, John J. and Eichenberger-Gilmore, Julie M. and Burns, Trudy L. and Stumbo, Phyllis J.},
title     = {{Dental caries and beverage consumption in young children.}},
journal   = {Pediatrics},
volume    = {112},
number    = {3 Pt 1},
year      = {2003}
}

@article{Levy2003,
author    = {Levy, S. M. and Warren, J. J. and Broffitt, B. and Hillis, S. L. and Kanellis, M. J.},
title     = {{Fluoride, beverages and dental caries in the primary dentition}},
journal   = {Caries Research},
volume    = {37},
number    = {3},
pages     = {157--165},
year      = {2003}
}

@article{Broffitt2013,
author    = {Broffitt, Barbara and Levy, Steven M. and Warren, John and Cavanaugh, Joseph E.},
title     = {{Factors associated with surface-level caries incidence in children aged 9 to 13: The Iowa Fluoride Study}},
journal   = {Journal of Public Health Dentistry},
volume    = {73},
number    = {4},
pages     = {304--310},
year      = {2013}
}

@article{Choo-Wosoba2018,
author    = {Choo-Wosoba, Hyoyoung and Gaskins, Jeremy and Levy, Steven and Datta, Somnath},
title     = {{A Bayesian approach for analyzing zero-inflated clustered count data with dispersion}},
journal   = {Statistics in Medicine},
volume    = {37},
number    = {5},
pages     = {801--812},
year      = {2018}
}

@article{Kang2023,
author    = {Kang, Tong and Gaskins, Jeremy and Levy, Steven and Datta, Somnath},
title     = {{Analyzing dental fluorosis data using a novel Bayesian model for clustered longitudinal ordinal outcomes with an inflated category}},
journal   = {Statistics in Medicine},
volume    = {42},
number    = {6},
pages     = {745--760},
year      = {2023}
}

@article{Bohning1999,
author    = {Dankmar B\"{o}hning and Ekkehart Dietz and Peter Schlattmann and Lisette Mendonca and Ursula Kirchner},
title     = {The Zero-Inflated Poisson Model and the Decayed, Missing and Filled Teeth Index in Dental Epidemiology},
journal   = {Journal of the Royal Statistical Society. Series A (Statistics in Society)},
publisher = {[Wiley, Royal Statistical Society]},
volume    = {162},
number    = {2},
pages     = {195--209},
year      = {1999}
}

@article{Sklar1959,
author    = {Sklar, M},
title     = {Fonctions de repartition an dimensions et leurs marges},
journal   = {Publ. inst. statist. univ. Paris},
volume    = {8},
pages     = {229--231},
year      = {1959}
}

@article{Kolev2009,
author    = {Kolev, Nikolai and Paiva, Delhi},
title     = {{Copula-based regression models: A survey}},
journal   = {Journal of Statistical Planning and Inference},
publisher = {Elsevier},
volume    = {139},
number    = {11},
pages     = {3847--3856},
year      = {2009}
}

@article{Smith2012,
author    = {Smith, Michael S. and Khaled, Mohamad A.},
title     = {{Estimation of copula models with discrete margins via Bayesian data augmentation}},
journal   = {Journal of the American Statistical Association},
volume    = {107},
number    = {497},
pages     = {290--303},
year      = {2012}
}

@article{Gardiner2009,
author    = {Gardiner, Joseph C. and Luo, Zhehui and Roman, Lee Anne},
title     = {{Fixed effects, random effects and GEE: What are the differences?}},
journal   = {Statistics in Medicine},
volume    = {28},
number    = {2},
pages     = {221--239},
year      = {2009}
}

@article{Zhang2012,
author    = {Zhang, H. and Yu, Q. and Feng, C. and Gunzler, D. and Wu, P. and Tu, X. M.},
title     = {{A new look at the difference between the GEE and the GLMM when modeling longitudinal count responses}},
journal   = {Journal of Applied Statistics},
volume    = {39},
number    = {9},
pages     = {2067--2079},
year      = {2012}
}

@article{Zeger1992,
author    = {Zeger, S. L. and Liang, K. Y.},
title     = {{An overview of methods for the analysis of longitudinal data}},
journal   = {Statistics in Medicine},
volume    = {11},
number    = {14-15},
pages     = {1825--1839},
year      = {1992}
}

@article{Lambert1992,
author    = {Diane Lambert},
title     = {Zero-Inflated Poisson Regression, with an Application to Defects in Manufacturing},
journal   = {Technometrics},
publisher = {[Taylor & Francis, Ltd., American Statistical Association, American Society for Quality]},
volume    = {34},
number    = {1},
pages     = {1--14},
year      = {1992}
}

@article{Rose2006,
author    = { C. E.   Rose  and  S. W.   Martin  and  K. A.   Wannemuehler  and  B. D.   Plikaytis },
title     = {On the Use of Zero-Inflated and Hurdle Models for Modeling Vaccine Adverse Event Count Data},
journal   = {Journal of Biopharmaceutical Statistics},
publisher = {Taylor & Francis},
volume    = {16},
number    = {4},
pages     = {463-481},
year      = {2006}
}

@article{Neuhaus1991,
author    = {Neuhaus, J. M. and Kalbfleisch, J. D. and Hauck, W. W.},
title     = {A {Comparison} of {Cluster}-{Specific} and {Population}-{Averaged} {Approaches} for {Analyzing} {Correlated} {Binary} {Data}},
journal   = {International Statistical Review / Revue Internationale de Statistique},
volume    = {59},
number    = {1},
pages     = {25},
year      = {1991}
}

@article{Li2002,
author    = {Y. Li and W. Wang},
title     = {Predicting Caries in Permanent Teeth from Caries in Primary Teeth: An Eight-year Cohort Study},
journal   = {Journal of Dental Research},
volume    = {81},
number    = {8},
pages     = {561-566},
year      = {2002}
}

@book{Daniels2008,
author    = {Daniels, M.J. and Hogan, J.W.},
title     = {Missing Data in Longitudinal Studies: Strategies for Bayesian Modeling and Sensitivity Analysis},
publisher = {Taylor \& Francis},
series    = {Chapman \& Hall/CRC Monographs on Statistics \& Applied Probability},
year      = {2008}
}

@book{Little2019,
author    = {Little, Roderick and Rubin, Donald},
title     = {Statistical {{Analysis}} with {{Missing Data}}},
publisher = {Wiley},
edition   = {3},
series    = {Wiley {{Series}} in {{Probability}} and {{Statistics}}},
month     = {April},
year      = {2019}
}

@manual{cremers2020waspr,
  title  = {{waspr}: Wasserstein Barycenters of Subset Posteriors},
  author = {Cremers, Jolien},
  year   = {2020},
  note   = {R package version 1.0.1},
  url    = {https://CRAN.R-project.org/package=waspr}
}

@article{Zafar2022,
  author  = {Zafar, Mahrukh and Levy, Steven M. and Warren, John J.
             and Xie, Xian-Jin and Kolker, Justine and Pendleton, Chandler},
  title   = {Prevalence of Non-Cavitated Lesions and Progression,
             Regression, and No Change from Age 9 to 23 Years},
  journal = {Journal of Public Health Dentistry},
  year    = {2022},
  volume  = {82},
  number  = {3},
  pages   = {313--320},
  pmid    = {35781658}
}

@article{Marshall2021,
  author  = {Marshall, Teresa A. and Curtis, Alexandra M.
             and Cavanaugh, Joseph E. and Warren, John J.
             and Levy, Steven M.},
  title   = {Beverage Intakes and Toothbrushing During Childhood
             Are Associated With Caries at Age 17 Years},
  journal = {Journal of the Academy of Nutrition and Dietetics},
  year    = {2021},
  volume  = {121},
  number  = {2},
  pages   = {253--260},
  pmid    = {33109505}
}

@article{hoffman2014no,
  title={The {No-U-Turn} sampler: adaptively setting path lengths in {Hamiltonian} {Monte Carlo}},
  author={Hoffman, Matthew D and Gelman, Andrew},
  journal={Journal of Machine Learning Research},
  volume={15},
  number={1},
  pages={1593--1623},
  year={2014}
}

@article{bingham2019pyro,
  title   = {Pyro: Deep Universal Probabilistic Programming},
  author  = {Bingham, Eli and Chen, Jonathan P. and Jankowiak, Martin and Obermeyer, Fritz and Pradhan, Neeraj and Karaletsos, Theofanis and Singh, Rohit and Szerlip, Paul A. and Horsfall, Paul and Goodman, Noah D.},
  journal = {Journal of Machine Learning Research},
  volume  = {20},
  pages   = {1--6},
  year    = {2019}
}

@article{mukherjee2024modeling,
  title={Modeling Zero-Inflated Correlated Dental Data through Gaussian Copulas and Approximate Bayesian Computation},
  author={Mukherjee, Anish and Gaskins, Jeremy T and Sarkar, Shoumi and Levy, Steven and Datta, Somnath},
  journal={arXiv preprint arXiv:2410.13949},
  year={2024}
}

@article{sarkar2024analyzing,
  title={Analyzing zero-inflated clustered longitudinal ordinal outcomes using GEE-type models with an application to dental fluorosis studies},
  author={Sarkar, Shoumi and Mukherjee, Anish and Gaskins, Jeremy T and Levy, Steven and Qiu, Peihua and Datta, Somnath},
  journal={arXiv preprint arXiv:2412.11348},
  year={2024}
}

@inproceedings{carvalho2009handling,
  title={Handling sparsity via the horseshoe},
  author={Carvalho, Carlos M and Polson, Nicholas G and Scott, James G},
  booktitle={Artificial intelligence and statistics},
  pages={73--80},
  year={2009},
  organization={PMLR}
}

\end{document}